\documentclass[a4paper,11pt]{article}
\pdfoutput=1 
\usepackage{graphicx}
\usepackage{float}
\usepackage{mciteplus}
\usepackage{dirtytalk}
\usepackage{slashed}
\usepackage{bbold}
\usepackage{braket}
\usepackage{mathtools,amsmath,amssymb,mathbbol}

\usepackage{jhep} 

\usepackage[T1]{fontenc} 

\renewcommand{\[}{\begin{equation}\begin{aligned}}
\renewcommand{\]}{\end{aligned}\end{equation}}
\def\beq{\begin{equation}}
\def\eeq{\end{equation}}

\renewcommand{\texttt}{{}}
\def\bs{\begin{subequations}}
\def\es{\end{subequations}}

\def\Cc{\mathcal{C}}

\def\Ec{\mathcal{E}}
\def\Fc{\mathcal{F}}

\def\Hc{\mathcal{H}}

\def\Lc{\mathcal{L}}

\def\Oc{\mathcal{O}}
\def\Pc{\mathcal{P}}

\def\Tc{\mathcal{T}}
\def\Uc{\mathcal{U}}
\def\Vc{\mathcal{V}}

\newcommand{\tia}[1]{}

\newcommand{\bea}{\begin{eqnarray}}
\newcommand{\eea}{\end{eqnarray}}
\newcommand{\beas}{\begin{eqnarray*}}
\newcommand{\eeas}{\end{eqnarray*}}
\newcommand{\bal}{\begin{aligned}}
\newcommand{\eal}{\end{aligned}}

\def\({\left(}
\def\){\right)}

\newcommand{\LF}{\left(}
\newcommand{\RF}{\right)}
\newcommand{\LT}{\left[}
\newcommand{\RT}{\right]}
\newcommand{\cpt}{$\Cc\Pc\Tc$}

\newcommand{\pd}{\partial}

\newcommand{\const}{\mathrm{const}}

\renewcommand{\imath}{\ensuremath{\mathrm{i}}}
\renewcommand{\vec}[1]{\ensuremath{\mathbf{#1}}}

\title{Towards a unitary formulation of quantum field theory in curved spacetime: the case of de Sitter spacetime}

\author[a]{K. Sravan Kumar}

\author[b]{Jo\~ao Marto}

\affiliation[a~]{Institute of Cosmology \& Gravitation,
	University of Portsmouth,
	Dennis Sciama Building, Burnaby Road,
	Portsmouth, PO1 3FX, United Kingdom}
\affiliation[b~]{
	Departamento de F\'isica, Centro de Matem\'atica e Aplicações (CMA-UBI), Universidade da Beira Interior, Rua Marquês D'Ávila e Bolama, 6201-001 Covilhã, Portugal }

\emailAdd{sravan.kumar@port.ac.uk}
\emailAdd{jmarto@ubi.pt}

\abstract{Before we ask what the quantum gravity theory is, it is a legitimate quest to formulate a robust quantum field theory in curved spacetime (QFTCS).
Several conceptual problems, especially unitarity loss (pure states evolving into mixed states), have raised concerns over several decades. 
In this paper, acknowledging the fact that {time} is a parameter in quantum theory, which is different from its status in the context of General Relativity (GR), we start with a "{quantum first approach}" and propose a new formulation for QFTCS based on the discrete spacetime transformations which offer a way to achieve unitarity. 
We rewrite the QFT in Minkowski spacetime with a direct-sum Fock space structure based on the discrete spacetime transformations and geometric superselection rules. Applying this framework to QFTCS, in the context of de Sitter (dS) spacetime, we elucidate how this approach to quantization complies with unitarity and the observer complementarity principle. We then comment on understanding the scattering of states in de Sitter spacetime. 
Furthermore, we discuss briefly the implications of our QFTCS approach to future research in quantum gravity.  }

\arxivnumber{arXiv:2305.06046}

\makeatother

\makeatletter
\gdef\@fpheader{}
\makeatother

\begin{document}
	
\maketitle

\section{Introduction}

Understanding of spacetime in the theory of relativity is dictated by light cones, geodesics, observers, and continuous coordinate transformations that are complemented by spacetime symmetries whether it is Lorentz invariance, Poincaré invariance or general covariance \cite{Wald:1984rg}. The most important aspects of quantum field theory (QFT) are S-matrix unitarity and the role of discrete spacetime transformations \cite{Coleman:2018mew,Roberts:2022xcj}. QFT in Minkowski spacetime is a merging of quantum mechanics (QM) and special relativity (SR). All QFT is built on the definition of a positive energy state (with respect to fixing an arrow of time) according to Schr\"{o}dinger equation and imitating the position and momentum commutation relations in QM  with the quantum field operator and its conjugate momenta. The unification of QM and SR is achieved by demanding vanishing commutators of the operators corresponding to space-like distances. The standard model (SM) of particle physics is developed from these foundations with additional elements of renormalizability and degrees of freedom corresponding to fundamental forces of nature such as strong, weak, and electromagnetic. 

Let us focus on the description of gravity given by General Relativity (GR). GR has been well tested over the years, and it is a (classical) theory showing an astonishing consistency, as the recent detection of gravitational waves from black hole mergers and the two beautiful observations of black hole shadows \cite{LIGOScientific:2016aoc,EventHorizonTelescope:2022xnr,Rummel:2019ads} illustrates. 
Apart from the success of GR as a classical theory, it poses three interrelated challenges.  The first challenge is modifying GR towards short-distance scales to eliminate its singular solutions, such as the Big Bang and black holes.  The main message of singular solutions in GR is that extreme curvature regimes or regions in the Universe require physics beyond GR. The second biggest challenge is how we can formulate a quantum theory of gravity that is unitary and renormalizable. This second challenge has been pursued by theoretical physicists over the past half-century, and there has been a lot of progress in this direction (see \cite{deBoer:2022zka,Buoninfante:2022ykf} for a brief review and references therein). Then, the 3rd biggest challenge is consistent QFT construction in curved spacetime. This challenge is very different from the first two, and it is rather about the (quantum) nature of gravity at scales that are not necessarily close to Planck scales. For example, {quantum} gravity effects at the black hole and early Universe cosmological (inflationary) horizons.  This challenge is often termed semi-classical physics, and its formulation has several ambiguities (See \cite{Giddings:2022jda}). Rigorously speaking, the third challenge seeks a consistent framework for defining quantum fields in a curved manifold. Although this subject has been researched extensively in recent decades, examples such as Hawking radiation and inflationary quantum fluctuations result from QFT in curved spacetime (QFTCS). However,  open questions still prevail about the unitarity, S-matrix formalism, uniqueness of vacuum, ambiguities associated with observers, resolution of the black hole information paradox, quantum to the classical transition of inflationary quantum fluctuations \cite{Giddings:2022ipt,Visser:1997gf,Mathur:2009hf,Calmet:2022swf,Almheiri:2012rt,BruceAllen,Kiefer:1998qe,Bousso:2004tv,Kiefer:2008ku,Perez:2005gh,Sudarsky:2009za}.  The quantization procedure in curved spacetime is a prerequisite that can ultimately help us in quests for the UV-complete theory of quantum gravity and spacetime singularity resolutions. One of the key points of this paper is to illustrate the understanding of spacetime from a QFT point of view, which we call {Quantum First Approach} (QFA). In this QFA, we clearly distinguish between the notions of {time} according to quantum mechanics (QM) and classical theory. We embrace the view that the thermodynamic meaning of {time} (which is associated with the second law of thermodynamics) emerges from a quantum theory only when we specify the initial and final states \cite{Hartle:2013tm,Hartle:2020his}.  Indeed, in quantum theory, an arrow of time always comes with our choice in specifying initial and final states, while the nature of quantum theory (without dynamical effects of gravity) is always time-symmetric \cite{tHooft:2018jeq}. The real question here is whether (quantum) laws of nature are time-symmetric in curved spacetime. A deeper question is how to define time reversal operation for quantum states in QFTCS without changing the nature of the universe that classical spacetime describes. 
It is generically believed that the discrete symmetries, such as \cpt\, must be broken in a dynamical spacetime \cite{Mavromatos:2003hr}. It may be totally natural to expect that \cpt\, symmetries are spontaneously broken in a dynamical spacetime. Still, it is important to consistently formulate it in quantum theory and possibly give a measure of \cpt\, violation if the spacetime is dynamical.\footnote{In this regard, our proposal of this paper is applied to inflationary quantum fluctuations that not only explained the CMB anomalies \cite{Gaztanaga:2024vtr,Gaztanaga:2024whs} (650 times better than standard inflation) but also derived parity asymmetry signatures for primordial gravitational waves \cite{Kumar:2022zff}.} A pertinent question is: if \cpt\ is bound to be violated in curved spacetime, how can one recover the  \cpt\ symmetry in the Minkowski limit, the short distance limit, or the local Lorentz limit? The answer cannot be just taking the metric of curved spacetime and imposing flat spacetime limit, rather the quest involves a deep understanding of quantum theory in dynamical spacetime. Since spacetime metrics are solutions of GR or any of its modifications with local Lorentz symmetry, \footnote{which means if we go to sufficiently short distance scales, we can neglect effects of spacetime curvature (far away from spacetime singularities)} there is a deeper relation between QFT in curved spacetime and taking a flat space limit. This is why the QFT in Rindler spacetime (that leads to the Unruh radiation \cite{Crispino:2007eb}) has similar unitarity issues (pure states evolving into mixed states) to the QFT in curved spacetime. If one has to solve the unitarity problem of QFTCS, it is unavoidable to relook at the historical developments of quantum theory and find the crucial links we have been missing. In a nutshell, this paper attempts to bring a new proposal for quantization that offers a new perspective. 

 In all of our successful endeavors in theoretical physics, we always build classical physics as a limit of quantum physics rather than the other way around. The prime reason we go from quantum to classical limit is that quantum physics is built on axioms that do not have classical analogs. The physical {intuition} that we have from classical physics consistently fails in quantum physics. The best examples are the quantum tunneling phenomenon, quantum entanglement, etc. It is practically impossible to enlist the number of instances where quantum theory has successfully surprised us, and it has been continuing to do so.\footnote{It is worth to note that the search for a deeper understanding and reformulation of QM is still a motivation for a very active field of research \cite{tHooft:2018jeq,tHooft:2020tuu}.}

Our current formulations of QFT in curved spacetime start with first defining observers with respect to a classical intuition about time, according to GR, and then a quantization in analogy with QFT in Minkowski spacetime \cite{Birrell:1982ix,Hollands:2009bke,Hollands:2014eia}. Moreover,  the choice of vacuum states leads to many questions and complexities related to observers, formulation of quantum states, and unitarity. Certainly, these questions cause numerous conceptual conundrums and are an obstacle to rightfully understanding quantization.  In several instances, curved spacetimes come with event or apparent horizons with causally disconnected regions of spacetime. In the context of de Sitter space (dS), Schr\"{o}dinger's conjecture \cite{Schrodinger1956}  demands that every observer should be able to see what is happening within his/her horizon using pure states, being the physics, that every observer perceives within the horizon, unitary.  This is later stated as the observer complementarity principle \cite{Parikh_2003}, which evades any information loss and leads to a consistent reconstruction of physics beyond the observer's horizon by looking at what is happening within the horizon.  In other words, Schr\"{o}dinger conjecture implies that one must be able to define a unitary QFT in curved spacetime consistent with the observer complementarity principle. We think the origin of many paradoxes in QFTCS is linked with the approach of first imagining an observer and defining coordinate patches with respect to the thermodynamical arrow of time before performing the quantization procedure. We instead decide to do the opposite with the QFA approach\footnote{Our "{quantum first approach}" wording is inspired from \cite{Giddings:2022jda}. }. 
  
 We proceed with our investigation by meticulously examining our understanding of the vacuum structure in Minkowski spacetime.  
 Then, we analyze the quantization issue in de Sitter spacetime realized in the flat Friedmann-Lema\^itre-Robertson-Walker (FLRW) metric. In the next step, we study the quantization in the closed FLRW Universe and briefly comment on constructing a unitary approach in the static de Sitter space context. Our study proposes a new picture for QFT in curved spacetime, preserving the unitarity. Thus, it differs from well-known approaches found in the literature  \cite{Birrell:1982ix,Ford:1997hb,Mukhanov:2007zz,Wald:1995yp}. Towards the end of the paper, we discuss differences between our approach and the standard quantization schemes in curved spacetime  \cite{Birrell:1982ix,Ford:1997hb}.  We also provide extensive reasoning for how our QFA approach opens doors for a new understanding of curved spacetime from the quantum point of view. 
 We organize our paper as follows.  
 
 

In \textbf{Sec.~\ref{sec:2to1}}, we discuss the de Sitter spacetime in FLRW coordinates and demonstrate the appearance of two arrows of time in realizing a description of one Universe. We present the viewpoints of how QFT in dS spacetime is historically carried with picking an arrow of time and treating the other Universe with the opposite arrow of time as a parallel Universe. We then define the path of our investigation of a quantum theory in one dS Universe with both arrows of time.

In \textbf{Sec.~\ref{sec:direct-sumQM}}, we present a direct-sum formulation of quantum mechanics with geometric superselection rules based on realizing $\Pc\Tc$ symmetry and defining a positive energy state with opposite arrows of time.
  
  {In \textbf{Sec.~\ref{sec:direct-sumFQFT}} we formulate a direct-sum QFT (DQFT) in Minkowski spacetime, which is a second-quantization of the direct-sum quantum mechanics established in Sec.~\ref{sec:direct-sumQM}.   }
 	
  In \textbf{Sec.~\ref{sec6DQFTDS}}, 
  We formulate DQFT in flat FLRW dS in analogy with the DQFT of Minkowski spacetime formulated in Sec.~\ref{sec:direct-sumFQFT}.  Then, we discuss how the unitarity and observer complementarity can be achieved with dS DQFT, which gives us an alternative view of the so far understanding of (quantum) dS spacetime that has been explored in various quantum gravity formulations over the years (See \cite{nlab:de_sitter_spacetime} and references therein).  Furthermore, we set a preliminary scattering problem in dS and define a unitary S-matrix, which, in the short distance limit, is reduced to the S-matrix of DQFT in Minkowski spacetime. Finally, we present the {conformal diagram} to represent dS DQFT in conjunction with comparisons to DQFT in Rindler spacetime explained in Appendix.~\ref{sec:Rindler}. 
 	
 	 In \textbf{Sec.~\ref{sec:DQFT-QG}}, We discuss in detail how our formulation of QFTCS can be further applicable in quantum gravity research. We especially discuss the difference between our construction and various understandings of dS spacetime inspired by string theory.
 	 
 	  In \textbf{Sec.~\ref{sec:conc}}, we provide a summary and conclusions with the future outlook of the QFTCS framework we initiated in this paper.

 Throughout this paper, we follow the conventions that $\hbar =c=1$, metric signature $(-,+,+,+)$ and the reduced Planck mass $M_p=\frac{1}{\sqrt{8\pi G}} =1$. 

 \section{One physical Universe with two arrows of time}
 \label{sec:2to1}

Quantum field theory in curved spacetime is challenged by two arrows of time that can describe the same physical Universe. We illustrate this with the realizations of de Sitter spacetime (dS) in FLRW coordinates. The dS manifold is characterized by its simple relations between the metric tensor and the curvature quantities, such as follows
\begin{equation}
R^{\mu}_{\nu\rho\sigma}  =\frac{R}{12}\LF \delta^\mu_\rho g_{\nu\sigma} -\delta^\mu_\sigma g_{\nu\rho}  \RF,\quad  R_{\nu\sigma} = \frac{R}{4}g_{\nu\sigma},\quad R=\const\,.
\label{dSdef}
\end{equation}
The above definition is coordinate independent, and, as it is well-known, dS spacetime is a solution to Einstein's GR with a cosmological constant. The dS spacetime in FLRW coordinates is most relevant for understanding early Universe cosmology. 

\subsection{De Sitter in flat FLRW coordinates}

Let us consider the dS metric in flat FLRW coordinates, which are widely used in the context of inflationary cosmology. 
\begin{equation}
	ds^2 = -dt^2 + a(t)^2d\textbf{x}^2
	\label{dsmetric}
\end{equation}
where the scale factor is given by
\begin{equation}
	a(t) = e^{Ht},\quad H^2 = \LF\frac{1}{a}\frac{da}{dt}\RF^2>0\,, 
\end{equation}
where $H$ here is the Hubble parameter. 
Each point in dS is surrounded by a horizon, given by the coordinate radius or also known as the comoving Hubble radius \cite{Riotto:2002yw}. 
\begin{equation}
	r_H = \Big\vert  \frac{1}{aH} \Big\vert
	\label{horizonds}
\end{equation}
To understand quantization, in dS spacetime, it is convenient to write dS metric \eqref{dsmetric} in terms of the conformal time defined by 
\begin{equation}
	d\tau = \frac{dt}{a(t)}\,. 
\end{equation}
Integrating this equation, we obtain
\[
\tau= 	-\frac{1}{a(t) H}
\,\text{,}
\]
Where the integration constant is set to zero.
In terms of this new time coordinate $\tau$, the dS metric is conformal to the Minkowski metric
\begin{equation}
	ds^2 = \frac{1}{H^2\tau^2}\LF -d\tau^2+ d\textbf{x}^2 \RF
	\label{conmetric}
\end{equation}
In terms of $\tau$, metric \eqref{conmetric} becomes conformal to Minkowski and provides a huge advantage for quantizing the KG field. We write the metric as \eqref{conmetric} to rewrite the effect of the curved spacetime as a harmonic oscillator with time-dependent mass. This is a convenient way to implement quantization. 

Note that the time coordinate $\tau \gtrless 0$ has a coordinate singularity at $\tau =0$. Nevertheless, it is just an artifact of the coordinate transformation; in reality, there is no spacetime singularity as such in de Sitter spacetime. We can clearly see that the dS metric satisfies $\Pc\Tc$ symmetries just like Minkowski spacetime 
\begin{equation}
	\Pc\Tc: \tau\to - \tau,\quad \textbf{x} \to -\textbf{x}\,. 
	\label{disflat}
\end{equation}
In quantum theory, we perform quantization and treat the conformal coordinate $\tau$ as a {time} parameter. We can notice that the time ($\tau$) reflection operation actually leads to time reversal in cosmic time, as well as the change of sign for the Hubble parameter $H$. 
\begin{equation}
\Tc:	\tau \to -\tau \implies t\to -t,\, H\to -H\,. 
\label{dSsym}
\end{equation}
It is really worth noting that $H\to -H$ operation does not change any dS curvature invariant. For example, the dS Ricci scalar $R=12H^2$ is completely symmetric under the change of sign in $H$. 

One simple observation we can make from \eqref{dsmetric}, is that we get an
\begin{equation}
	{\rm Expanding\,Universe:} \implies \begin{cases}
		t: -\infty \to +\infty,\quad  H>0 & (\tau: -\infty \to 0)\\ 
		t: +\infty \to -\infty,\quad  H<0 & (\tau: \infty \to 0)
	\end{cases}
\label{expcon}
\end{equation}
and a 
\begin{equation}
	{\rm Contracting\,Universe:} \implies \begin{cases}
		t: +\infty \to -\infty,\quad  H>0 & (\tau: 0\to -\infty )\\ 
		t: -\infty \to +\infty,\quad  H<0 & (\tau: 0\to \infty)
	\end{cases}
\label{contrcon}
\end{equation}
Expansion and contraction of the universe are now determined by the conformal time  $\tau$ flow:
\[
\text{Expansion: } \tau: \mp \infty \longrightarrow 0\,\, (r_H \,\, {\rm shrinks})\,\text{,}&\qquad& \text{Contraction: } 0 \longrightarrow \mp \infty\, \, (r_H \,\, {\rm grows})\,,
\]
In the literature, it is often assumed that $H>0$ and $\tau<0$ indicates an expanding Universe \cite{Bousso:2002fq,Kim:2002uz,Spradlin:2001pw,Hartman:2017}. {All correlation functions in dS space are computed using this assumption \cite{Jain:2022uja,Rajaraman:2015dta,DiPietro:2021sjt}.   }
Notice that the time reversal operation does not change the nature of the Universe, i.e., an expanding Universe remains expanding, and a contracting Universe remains contracting (\eqref{expcon} and \eqref{contrcon}). In the literature, it is generally assumed that these two expanding branches (\eqref{expcon}) (similarly, two contracting branches \eqref{contrcon}) are two expanding (similarly, two contracting) Universes. We adopt the notion that an expanding Universe is defined by its shrinking, comoving horizon (growth of scalar factor which acts as a clock), and the time reversal operation by \eqref{expcon} does not change the character of the expanding Universe.  It is this time reversal operation that does not change the nature of the Universe; it is vital for the construction of a consistent QFTCS.\footnote{We come across a recent paper \cite{Donath:2024utn} that considers $\tau<0$ and $\tau>0$ for the construction of in-out correlators in QFT in dS spacetime, however, in their work $\tau>0$ is treated as contracting Universe branch. In contrast, in our case, we define a time reversal operation ($\tau\to -\tau$ that goes with $H\to -H$ and $t\to -t$) in the expanding Universe.} 

\subsection{De Sitter in closed FLRW coordinates}

Let us consider dS space expressed in the closed FLRW coordinates, also known as the global coordinates \cite{Kim:2002uz,Mukhanov:2005sc}. 
\begin{equation}
ds^2 = -dt^2+ \cosh^2\LF Ht \RF \LT d\chi^2+ \sin^2\chi d\Omega^2 \RT \,,
\label{cloFLRW}
\end{equation}
where $0\leq \chi = \frac{\sin^{-1}\LF \sqrt{6} Hr \RF}{\sqrt{6} H}\leq \pi$. 
The above metric Ricci scalar is $R=12H^2$. This metric describes the contraction and expansion of a 3-dimensional sphere. For quantization purposes, it is useful to write metric \eqref{cloFLRW} by defining conformal time as
\begin{equation}
	\tau =  \int \frac{dt}{a(t)} = \frac{2}{H}\tan ^{-1}\left(\tanh \left(\frac{H t}{2}\right)\right)\,. 
	\label{concFLRW}
\end{equation}
The metric, in terms of this conformal time, takes the form 
\begin{equation}
	ds^2 = \frac{1}{\cos^2\tau} \LT -d\tau^2 + d\chi^2+\sin^2\chi d\Omega^2\RT
	\label{conmetrc}
\end{equation}
The metric \eqref{conmetrc} discrete symmetries are 
\begin{equation}
	\Pc\Tc: \tau \to -\tau,\quad \chi\to \chi,\quad \LF \theta,\,\varphi \RF\to \LF \pi-\theta,\,\varphi+\pi \RF\,. 
	\label{discflrw}
\end{equation}
As for flat FLRW dS, the time reversal operation involves flipping the sign of $H$. This means 
\begin{equation}
	{\rm contraction\,to\,expansion} \implies \begin{cases}
		t: -\infty \to +\infty, \quad H>0,\quad \LF \tau: -\frac{\pi}{2} \to \frac{\pi}{2} \RF \\ 
			t: +\infty \to -\infty, \quad H<0,\quad \LF \tau: \frac{\pi}{2} \to -\frac{\pi}{2} \RF
	\end{cases}
\label{clldS}
\end{equation}
Therefore, even in dS global coordinates, the sign of $H$ plays a crucial role in the QFT formulation. We emphasize that the time reversal operation, considered here, is a purely quantum mechanical concept with no classical analog (See \cite{Rovelli:2000aw,Rovelli:2004tv} for more discussions about the quantum mechanical concept of time). Therefore, \eqref{clldS} represents one Universe evolving from contraction to expansion; the discrete transformation applied from the first to the second line of \eqref{clldS} is exactly what we call time reversal operation in closed dS.

\subsection{De Sitter in static coordinates}

De sitter spacetime in the static coordinates $\LF t_s,\,r \RF$ can be expressed as  \cite{Griffiths:2009dfa,nlab:de_sitter_spacetime}
 \begin{equation}
\begin{aligned}
    ds^2 & = -\LF 1-H^2r^2 \RF dt_s^2 + \frac{1}{\LF 1-H^2r^2 \RF}dr^2 + r^2d\Omega^2 \\ 
    &  = \frac{1}{H^2\LF 1-\Tilde{\Uc}\Tilde{\Vc} \RF^2} \LF -4d\Tilde{\Uc}d\Tilde{\Vc}+ \LF 1+\Tilde{\Uc}\Tilde{\Vc} \RF^2 d\Omega^2 \RF
    \end{aligned}
    \label{statdS}
\end{equation}
where $r= \big\vert \frac{1}{H}\big\vert \frac{1+\Tilde{\Uc}\Tilde{\Vc}}{1-\Tilde{\Uc}\Tilde{\Vc}}$ and $\frac{\Tilde{\Vc}}{\Tilde{\Uc}} = -e^{2Ht_s}$.
The discrete symmetries, in this case, are 
\begin{equation}
	\Pc\Tc: t_s \to -t_s,\,r\to r,\quad \quad \LF \theta,\,\varphi \RF\to \LF \pi-\theta,\,\varphi+\pi \RF\,. 
	\label{statdiscr}
\end{equation}
together with another symmetry, i.e., $H\to -H$. The static de Sitter \eqref{statdS} shares many similarities with the black hole case in the context of horizon quantum physics and thermodynamics \cite{Gibbons:1977mu}.

The relation between the flat FLRW dS \eqref{dsmetric} and \eqref{statdS} can be worked out as 
\begin{equation}
    r=r_fe^{Ht},\quad e^{-2Ht_s} = e^{-2Ht}-H^2r_f^2\,,
\end{equation}
where $r_f$ is the radial coordinate of flat FLRW dS \eqref{dsmetric} and the angular coordinates $\Omega \equiv \LF \theta,\,\varphi \RF$ remain the same in both metrics.

Similarly, we can draw the relation between \eqref{cloFLRW} and \eqref{statdS} by the following coordinate transformations 
\begin{equation}
  r= \frac{1}{H}\cosh\LF Ht \RF \sin \chi\,,\quad \sinh^2\LF Ht_s \RF = \frac{\sinh^2\LF Ht \RF}{1-r^2H^2}\,,
\end{equation}

\subsection{Foundational questions on two arrows of time}

Thus, from the above sections, we can conclude that there are two arrows of time that represent the same physical universe. Choosing an arrow time by hand before doing a quantization breaks the symmetry of the spacetime. Schr\"{o}dinger in 1956 proposed to identify parity conjugate regions with opposite arrows of time in dS spacetime to represent one physical expanding Universe \cite{Schrodinger1956}. As stated in \cite{Schrodinger1956} 
\begin{quote}
    An event in de Sitter spacetime has to be described by thin rods connecting the antipodal points in spacetime $\LF \tau,\,\textbf{x} \RF$ and $\LF -\tau,\, -\textbf{x} \RF$
\end{quote}
The above understanding of Schr\"{o}dinger's is in a quantum mechanical sense, and it basically demands an understanding of quantum theory with two arrows of time. We will later see if such a construction is possible, which will lead us to a unitary description of curved spacetime. 

Alternatively, what is usually taken in literature is picking an arrow of time out of the two possibilities \eqref{expcon}. If we do that, we first break the symmetry by our choice, and secondly, we must admit there is another Universe with the opposite arrow of time or whatever the other Universe is; we discard it as unphysical. This deduction is entirely observer-dependent. The majority of the research in literature \cite{Hartman:2020khs,Shaghoulian:2021cef,Shaghoulian:2022fop,Balasubramanian:2001rb,Balasubramanian:2021wgd,Bousso:2002fq,Colas:2024xjy} has pondered around the thought of causally disconnected entangled quantum Universes where unitarity (pure states evolving into pure states) is lost for an observer living in either of the Universe. This approach has led to the construction of so-called thermofield double states in the dual Minkowski and dual FLRW spacetimes \cite{Hartman:2020khs}. 

Our study is entirely different from all these; we focus on formulating quantum theory with two arrows of time in one Universe with unitarity and observer complementarity. This line of thought requires us to understand quantum theory in a new way as we will discuss in the next sections.  

 \section{Direct-sum Hilbert space construction of quantum mechanics}

\label{sec:direct-sumQM}

This section proposes a new view of quantum mechanics (QM) using geometric superselection sectors based on parity and time reversal operations. The discussion here is the crux of our proposal for the unitary formulation of QFTCS in the later sections. We present here a direct-sum formulation of QM, which explicitly represents the $\Pc\Tc$ symmetry of physical systems.  

 Historically, QFT formulation (second quantization) has been hailed with significant insights from QM. It has been noted by several studies that QM is time symmetric \cite{tHooft:2018jeq,Donoghue:2019ecz}
 Recent studies, both in the theoretical and experimental domains, are worth noticing about the time-symmetric nature of quantum theory \cite{geiger2019theory,Stromberg:2022dgp,Harshman}. However, both in QFT and in QM, one assume an arrow of time with respect to which a positive energy state is defined. 

Let us recall the Schr\"{o}dinger equation
\begin{equation}
	i\frac{\pd \vert \Psi \rangle}{\pd t_p} = \hat{H}\vert \Psi \rangle
	\label{sch1}
\end{equation}
where $\hat{H}$ is the Hamiltonian, whose precise form is irrelevant to the present discussion. However, we assume it to be $\Pc\Tc$ symmetric as many physical systems (without involving gravity) in nature are.\footnote{Because of the fact that \eqref{HPTsym} holds well in many instances, there is a whole field of research on $\Pc\Tc$ symmetric formulations of QM (See \cite{Bender:2005tb} for more details). However, our framework does not fall into the so-called $\Pc\Tc$ symmetric QM of \cite{Bender:2005tb}. } Thus, we have
\begin{equation}
	[\hat{H},\, \Pc\Tc] =0\,. 
	\label{HPTsym}
\end{equation}
Following \eqref{sch1}, usually one defines a positive energy state 
\begin{equation}
\vert \Psi\rangle_{t_p} = e^{-i\Ec t_p}\vert \Psi\rangle_0,\quad t_p:-\infty\to \infty 
\label{actchoice}
\end{equation}
with respect to a convention on the arrow of time. Here, $\Ec$ is the energy eigenvalue of the (time-independent) Hamiltonian $\hat{H}$.

\begin{figure}
    \centering
    \includegraphics[width=0.5\linewidth]{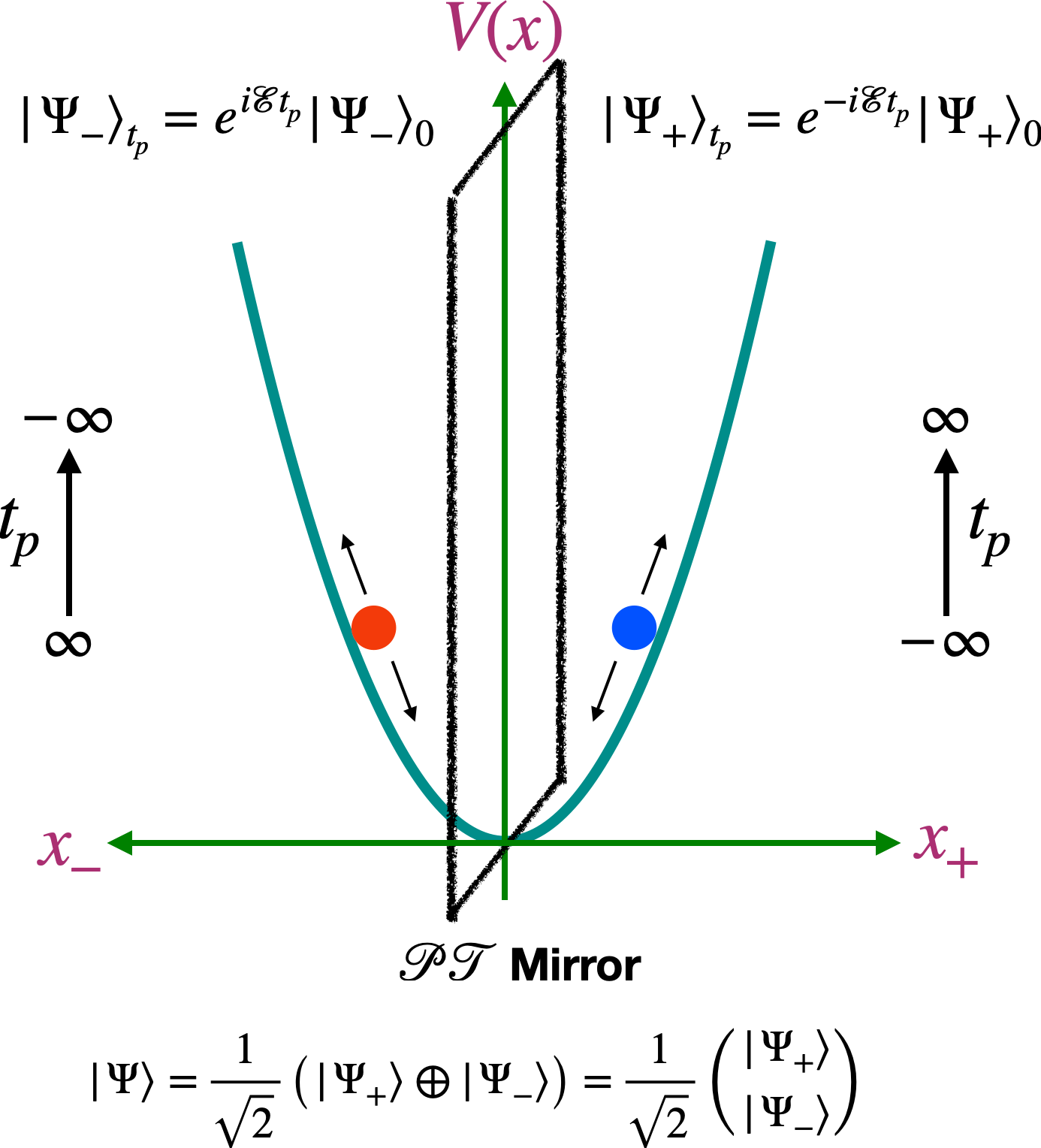}
    \caption{This figure depicts the understanding of a quantum harmonic oscillator with geometric superselection Hilbert spaces attached to parity conjugate regions to describe a single component of the quantum state as a direct sum of two components. The components state $\vert  Psi_\rangle $ in this picture share the positive and negative position eigenvalues and evolve with opposite arrows of time in the parity conjugate regions. It is like one places a "PT mirror" in the middle; the evolution of component states is restricted to their respective regions complimenting each other.  }
    \label{fig:hodq}
\end{figure}

Let us now take the Schr\"odinger equation \eqref{sch1} but by replacing $i\to -i$ 
\begin{equation}
	-i\frac{\pd\vert \Psi\rangle }{\pd t_p} = \hat{H}\vert \Psi\rangle
	\label{sch2}
\end{equation}
The state vector, according to \eqref{sch2}, evolves as 
\begin{equation}
    \vert \Psi\rangle_{t_p} = e^{i\Ec t_p}\vert \Psi\rangle_0,\quad t_p:\infty \to -\infty
    \label{altpositve}
\end{equation}
and it is the positive energy state but with the opposite convention to the arrow of time. So, the arrow of time is associated with whether we have $+i$ or $-i$ in quantum theory, as explained well in \cite{Donoghue:2019ecz}.
 
 Here, we attempt to make a new formulation of QM built with both arrows of time. Recalling again that time is a parameter in quantum theory\footnote{In quantum theory, we talk about position operators, not the time operators, and this is attributed to the anti-unitary character of time reversal operation found by Eugine Wigner in 1931 \cite{Roberts:2022xcj}.}, 
\begin{quote}
	{\it We express position space operators (and associated state vectors) as direct-sum of two components based on the parity operation  and assign opposite arrows of time to the parity conjugate regions. Thus, the~total Hilbert space $\Hc$ of a single particle state becomes the direct-sum of the sectorial Hilbert spaces representing parity conjugate regions.}
\end{quote}
Therefore, in the direct-sum Hilbert space construction of QM, the single-particle quantum state is a two-component state
\begin{equation}
	\vert \Psi\rangle = \frac{\vert \Psi_+\rangle \oplus \vert \Psi_{-}\rangle }{\sqrt{2}} = \frac{1}{\sqrt{2}} \begin{pmatrix}
		\vert \Psi_+\rangle & \\ \vert \Psi_{-}\rangle 
	\end{pmatrix}\,. 
	\label{disumn}
\end{equation}
The factor of $\sqrt{2}$ in \eqref{disumn} is a normalization to have the total probabilities add up to 1 as 
\begin{equation}
\int_{-\infty}^\infty dx\langle \Psi \vert \Psi \rangle = \int^{\infty}_0 dx_+\frac{\langle \Psi_+ \vert \Psi_+\rangle }{2}+\int^{0}_{-\infty} dx_{-}\frac{\langle \Psi_{-}\vert \Psi_{-}\rangle }{2}  =1\,,
\end{equation}
where (1D) parity conjugate position coordinates $x_+ = x\gtrsim 0$ whereas $x_{-} = x\lesssim 0$. 

This clearly shows the concept of direct-sum, which is totally different from the usual summation (i.e., the superposition of states) in quantum theory. In \eqref{disumn}, state vector $\vert \Psi\rangle$ is defined in a Hilbert space, which can be written as 
\begin{equation}
	\Hc = \Hc_+\oplus \Hc_{-}\,, 
	\label{totH}
\end{equation}
where $\Hc_+$ and $\Hc_{-}$ are the Hilbert spaces corresponding to the regions, which are parity conjugates of each other. Since the states in $\Hc_\pm$ cannot form a superposition and they share regions of physical space related by parity, we call $\Hc_\pm$ as geometric superselection sectors. This concept is distinct from the superselection sectors known in algebraic QFT \cite{nlab:superselection_theory}.\footnote{Superselection rules in QFT were first formulated by  Wick, Wightman, and Wigner in the context of intrinsic parity of elementary particles \cite{Wick:1952nb,nlab:superselection_theory,Giulini:2007fn}. The main idea of this is that a Hilbert space can be written as a direct-sum of subspaces, called superselection sectors, which completely forbids the existence of any superposition of state vectors ailing from different superselection sectors.}
QM in this framework is governed by the direct-sum Schr\"{o}dinger equation
\begin{equation}
	i\frac{\pd}{\pd t_p} \begin{pmatrix}
		\vert \Psi_+\rangle & \\ \vert \Psi_{-}\rangle 
	\end{pmatrix} = \begin{pmatrix}
		\hat{H}_+ & 0 \\ 
		0 & 	-\hat{H}_- \end{pmatrix}\begin{pmatrix}
		\vert \Psi_+\rangle & \\ \vert \Psi_{-}\rangle
	\end{pmatrix}\,, 
\label{sch3}
\end{equation}
where $\hat H_\pm \LF \hat x_\pm\oplus \hat p_\pm \RF$ are two components of the full Hamiltonian ($\hat H=\hat H_+\oplus H_-$), which are functions of position and momentum operators of parity conjugate regions 
whose canonical commutations are given by
\begin{equation}
	[\hat{x}_+,\,\hat{p}_+] = i\hbar\quad \hat{p}_+= -i\hbar \frac{\pd}{\pd x_+}\,,
\end{equation}
and
\begin{equation}
	[\hat{x}_-,\,\hat{p}_-] = -i\hbar\quad \hat{p}_-= i\hbar \frac{\pd}{\pd x_-}\,,
\end{equation}
where the change of $i\to -i$ is a consequence of the time reversal operation. Furthermore, note that 
\begin{equation}
    \Big[ \hat x_+,\,\hat x_- \Big]= \Big[ \hat p_+,\,\hat p_- \Big]= \Big[ \hat x_+,\,\hat p_- \Big] = \Big[ \hat p_+,\,\hat x_- \Big] =0\,.
\end{equation}
This is because operators from the geometric superselection sectors commute with each other. This rule also ascertains that physics at parity conjugate points with opposite arrows of time cannot affect each other. This is a way to avoid closed timelike curves. Note that here, the position operators $\hat x_\pm$ have parity conjugate eigenvalues $x_\pm$.

Following \eqref{totH}, the operators (say $\hat{\Oc}$) that act on the state vectors in the total Hilbert space become the direct-sum of the operators corresponding to the sectorial Hilbert space
\begin{equation}
\hat{	\Oc} = \hat{\Oc}_+\oplus \hat{\Oc}_{-} = \begin{pmatrix}
	\hat{\Oc}_+ & 0 \\ 
	0 & 	\hat{\Oc}_{-}
\end{pmatrix}
\label{disumOp}
\end{equation}
As we see in \eqref{disumOp}, the direct-sum of operators means extending the matrix operators into block diagonal form, which acts on the higher dimensional state vectors $\vert \Psi\rangle$ as written in \eqref{disumn}.

It is important to note that 
\begin{equation}
\begin{aligned}
       \hat{H}\vert \Psi_{+} \rangle & = \LF \hat{H}_+ \oplus \hat{H}_{-} \RF \vert \Psi_+\rangle = \hat{H}_{+}\vert \Psi_{+} \rangle \\ 
           \hat{H}\vert \Psi_{-} \rangle & = \LF \hat{H}_+ \oplus \hat{H}_{-} \RF \vert \Psi_-\rangle = \hat{H}_{-}\vert \Psi_{-} \rangle 
    \label{opact} 
\end{aligned}
\end{equation}
This means operators of Hilbert space $\Hc_{+}$ do not act on the states of $\Hc_{-}$ and vice versa. 

In direct-sum QM, the wave function of the quantum harmonic oscillator would become \cite{Gaztanaga:2024vtr,Kumar:2024quv} 
\begin{equation}
   \Psi_n(x) = \langle x \vert \Psi_n\rangle \equiv \begin{cases}
       \frac{1}{\sqrt{2^{n+1}n!}}\LF \frac{1}{\pi} \RF^{1/4} e^{-\frac{1}{2}x_+^2} H_n\LF x_+ \RF e^{-iE_n t_p},\quad x_+\gtrsim 0\\ 
      \frac{1}{\sqrt{2^{n+1}n!}}\LF \frac{1}{\pi} \RF^{1/4} e^{-\frac{1}{2}x_-^2} H_n\LF x_- \RF e^{iE_n t_p},\quad x_- \lesssim 0
   \end{cases}
   \label{hopsi}
\end{equation}
where $E_n$ is energy for each $n$ of the Hermite polynomials $H_n$. In the limit $x_\pm \to 0_\pm$, the two-component states $\vert \Psi_{\pm}\rangle$ match automatically. The point $x=0$ is not a special point, we can apply any continuous coordinate transformations the $\Pc\Tc$ operations remain the same. Thus, we can always associate our geometric superselection sectors $\Hc_\pm$ with parity conjugate regions of physical space. 
Finally, the states \eqref{hopsi} corresponding to different energies that satisfy the orthogonality relation
 \begin{equation}
     \int_{-\infty}^{\infty} \langle \Psi_n\vert \Psi_m\rangle dx = \int_0^\infty dx_+ \langle \Psi_{n+}\vert \Psi_{m+}\rangle + \int_{-\infty}^0 dx_- \langle \Psi_{n-}\vert \Psi_{m-}\rangle dx_- = \delta_{n,m}
 \end{equation}
Note that the arrow of time in quantum theory is not observable, and it is not associated with any classical notion of time. The schematic description of the quantum harmonic oscillator in direct-sum quantum theory is depicted in Figure.~\ref{fig:hodq}.

\section{Direct-sum quantum field theory (DQFT) in Minkowski spacetime} 

\label{sec:direct-sumFQFT}

QFT in Minkowski spacetime combines QM with special relativity by demanding that operators at spacelike distances must commute. We rely on the definition of positive energy state to establish quantum fields in Minkowski spacetime. However, the Minkowski metric
\begin{equation}
    ds^2 = -dt_p^2+d\textbf{x}^2
    \label{min}
\end{equation}
is invariant under $\Pc: \textbf{x}\to -\textbf{x}$ and $\Tc: t_p\to -t_p$. Thus, if we choose an arrow of time for a positive energy state (as in \eqref{actchoice}), we implicitly abandon the alternative choice \eqref{altpositve}. The direct-sum QM discussed in the previous section presents a formalism that includes both arrows of time in the single description of the quantum state, which we elevate to quantum fields in Minkowski spacetime. 

Now, let us consider the case of Klein-Gordian (KG) field quantization in DQFT. We propose that the KG field operator is a direct-sum of the two components 
\begin{equation}
	\begin{aligned}
	\hat{\phi}\LF x \RF & = \frac{1}{\sqrt{2}}  \hat{\phi}_{+}  \LF t_p,\, \textbf{x} \RF \oplus \frac{1}{\sqrt{2}} \hat{\phi}_{-} \LF -t_p,\,-\textbf{x} \RF \\ 
	& = \frac{1}{\sqrt{2}} \begin{pmatrix}
		\hat{\phi}_{+} & 0 \\ 
		0 & 	\hat{\phi}_{-}
	\end{pmatrix}
	\end{aligned}
	\label{disum}
\end{equation}
where 
\begin{equation}
	\begin{aligned}
	 \hat{\phi}_{+}  &  = 	\int \frac{d^3\vec{k}}{\LF 2\pi\RF ^{3/2}	}	\frac{1}{\sqrt{2 \omega_\vec{k}}} \Bigg[\hat a_{+\,\vec{k}}  e^{ik\cdot x}+\hat a^\dagger_{+\,\textbf{k}} e^{-ik\cdot x} \Bigg] \\ 
	 	 \hat{\phi}_{-}   &  = 	\int \frac{d^3\vec{k}}{\LF 2\pi\RF ^{3/2}	}	\frac{1}{\sqrt{2 \omega_\vec{k}}} \Bigg[\hat a_{-\,\vec{k}}  e^{-ik\cdot x}+\hat a^\dagger_{-\,\textbf{k}} e^{ik\cdot x} \Bigg]\,, 
	\end{aligned}
\label{fiedDQFTMin}
\end{equation}
where the operators $a_+,\, a_+^\dagger$ and  $a_{-},\, a_{-}^\dagger$ satisfy the canonical commutation relations, and all of their mixed commutators are zero. 
\begin{equation}
	\begin{aligned}
		[\hat{a}_{+\,\textbf{k}},\,\hat{a}_{+\,\textbf{k}^\prime}^\dagger] & = 	[\hat{a}_{-\,\textbf{k}},\,\hat{a}_{-\,\textbf{k}^\prime}^\dagger] = \delta^{(3)}\LF \textbf{k}-\textbf{k}^\prime \RF\\
			[\hat{a}_{+\,\textbf{k}},\,\hat{a}_{-\,\textbf{k}^\prime}] &=	[\hat{a}_{+\,\textbf{k}},\,\hat{a}_{-\,\textbf{k}^\prime}^\dagger] = [\hat{a}_{+\,\textbf{k}}^\dagger,\,\hat{a}_{-\,\textbf{k}^\prime}^\dagger]  =0\,.
		\end{aligned}
	\label{comcan}
\end{equation}
The mixed correlation relations (i.e., the second line in \eqref{comcan}) must be zero to respect locality and causality in the sense that the operators corresponding to forward in time at position $\textbf{x}$ should commute with operators corresponding to backward in time at position $-\textbf{x}$. Furthermore, the commutation relations of the field and the corresponding conjugate momenta become 
\begin{equation}
	[\hat{	\phi}_+\LF t_p, \textbf{x} \RF,\,\pi_+\LF t_p,\,\textbf{x}^\prime \RF] = i \delta\LF \textbf{x}-\textbf{x}^\prime \RF,\quad [\hat{	\phi}_{-}\LF- t_p, -\textbf{x} \RF,\,\pi_{-}\LF -t_p,\,-\textbf{x}^\prime \RF] = -i \delta\LF \textbf{x}-\textbf{x}^\prime \RF,
	\label{canDQFT}
\end{equation}
where 
\begin{equation}
	\pi_I \LF t_p,\,\textbf{x} \RF = \frac{\pd\Lc_{\rm KG}}{\pd\LF \pd_{t_p}  \phi_+\RF},\quad  	\pi_{-} \LF t_p,\,\textbf{x} \RF = - \frac{\pd\Lc_{\rm KG}}{\pd\LF \pd_{t_p}  \phi_{-}\RF}
\end{equation}
We now define Fock space vacuums as 
\begin{equation}
	\hat{a}_{+\,\textbf{k}}\vert 0\rangle_+ = 0,\quad 	\hat{a}_{-\,\textbf{k}}\vert 0\rangle_{-} = 0 \, ,
\end{equation}
and the total Fock space vacuum state is given by 
\begin{equation}
	\vert 0\rangle_T = \vert 0_+\rangle \oplus \vert 0_-\rangle =  \begin{pmatrix}
		\vert 0_+\rangle \\ \vert 0_-\rangle\
	\end{pmatrix}\,.
	\label{tFs}
\end{equation}
First, it is important to note that one must not confuse direct-sum {$ \oplus $}  with the usual summation  {$ +$}. The direct sum operation has a very special meaning in mathematics \cite{Conway,Harshman,Mazenc:2019ety}, for example, a direct-sum of two matrices with dimension $m$ and $n$ becomes $m+n$ (which is different from the direct product $\otimes$ where the dimensionality becomes $mn$). The field operator, defined as the direct-sum of two operators, is not the same as under the usual summation of two parts. The latter operation is trivial, but the former is non-trivial. 
The physical meaning of  \eqref{disum} is the following. In QFT, field operator $\hat{\phi}\LF x \RF$ is a function of space $-\infty\leq \textbf{x}\leq\infty $ and time is a parameter with the convention $t_p: -\infty \to \infty$. In our approach, we split it as the direct-sum of two operators, where each is a function of parity conjugate points with opposite arrows of time. 
We can anticipate that this splitting changes the KG field operator structure and highlights the discrete spacetime transformations $\Pc\Tc$ that leave the Minkowski metric invariant. In our representation, the field operator acting on the total vacuum $\hat{	\phi}\LF t,\,\textbf{x} \RF \vert 0\rangle_T$ enables to create a positive energy state at position $\textbf{x}$ (by $\hat{\phi}_+\vert 0_+\rangle$) and at the same moment creates a positive energy state ($\hat{\phi}_-\vert 0_-\rangle$) at $-\textbf{x}$. Note that the field operator, defined through a direct-sum, represents only a single degree of freedom. The two Fock spaces $\Fc_\pm$ are geometric superselection sectors associated with the discrete spacetime transformations $\Pc\Tc$, and the total Fock space is the one that describes the quantum fields in the Minkowski spacetime vacuum. 
We emphasize that there is only one degree of freedom and not two. The two Fock spaces $\Fc_\pm$ are complementary; one cannot exist without the other. 
{The structure of total Fock space in DQFT in terms of Hilbert spaces of multiparticle states looks like
	\begin{equation}
		\Fc_T = \Fc_+ \oplus \Fc_{-}\,,
		\label{disumFock}
	\end{equation}
where 
\begin{equation}
	\begin{aligned}
	\Fc_+\LF \Hc \RF&  = \bigoplus_{i=0}^{\infty} \Hc_+^i = \Cc\oplus \Hc_{+1}\oplus \LF \Hc_{+1}\otimes \Hc_{+2}\RF\oplus \LF \Hc_{+1}\otimes\Hc_{+2}\otimes \Hc_{+3} \RF\oplus\cdots \\
		\Fc_{-}\LF \Hc \RF & = \bigoplus_{i=0}^{\infty} \Hc_{-}^i = \Cc\oplus \Hc_{-1}\oplus \LF \Hc_{-1}\otimes \Hc_{-2}\RF\oplus \LF \Hc_{-1}\otimes\Hc_{-2}\otimes \Hc_{-3} \RF\oplus\cdots\,, 
		\end{aligned}
\end{equation}
where $\Hc_{+n}$ and $\Hc_{-n}$ are n$^{\rm th}$ particle geometric superselection sector Hilbert spaces.
}

We note that  correlation functions, in this Fock space direct-sum,  can be computed as
\begin{equation}\begin{aligned}
		{}_{T}\langle 0 \vert \hat{	\phi}\LF x \RF \hat{	\phi}\LF x^\prime \RF\vert 0\rangle_{T} &= 	\frac{1}{2}\langle 0_+ \vert \hat{	\phi}_+\LF x \RF \hat{	\phi}_+\LF x^\prime \RF\vert 0_+\rangle + \frac{1}{2}\langle 0_- \vert \hat{	\phi}_{-}\LF -x \RF \hat{	\phi}_{II}\LF -x^\prime \RF\vert 0_-\rangle \, ,
\end{aligned}\end{equation}
Since the Minkowski spacetime is $\Pc\Tc$ symmetric, we get 
\begin{equation}
\langle 0_+ \vert \hat{	\phi}_+\LF x \RF \hat{	\phi}_+\LF x^\prime \RF\vert 0_+\rangle = \langle 0_- \vert \hat{	\phi}_{-}\LF x \RF \hat{	\phi}_{-}\LF x^\prime \RF\vert 0_-\rangle
\label{CPTnewdef}
\end{equation}
The Feynmann propagator is
\begin{equation}\begin{aligned}
		{}_{T}\langle 0 \vert \bar T \hat{	\phi}\LF x \RF \hat{	\phi}\LF x^\prime \RF\vert 0\rangle_{T} &= 	\frac{1}{2}\langle 0_+ \vert \bar T\hat{	\phi}_+\LF x \RF \hat{	\phi}_+\LF x^\prime \RF\vert 0_+\rangle + \frac{1}{2}\langle 0_- \vert \bar T\hat{	\phi}_{-}\LF -x \RF \hat{	\phi}_{II}\LF -x^\prime \RF\vert 0_-\rangle \, ,
\end{aligned}\end{equation}
where $\bar T$ is the ordered product. This means the Feynmann propagator has two parts each describing the field propagation in the parity conjugate regions associated with $\hat \phi_{\pm}$. The Feynmann propagator for $\hat \phi_{\pm}$ in Momentum space can be realized through 
\begin{equation}
	-i\Delta_F^{+}\LF x-y \RF = -i\int \frac{d^4k}{\LF 2\pi \RF^4}\frac{ 1}{k^2+m^2-i\epsilon},\quad 	i\Delta_F^{-}\LF x-y \RF = i\int \frac{d^4k}{\LF 2\pi \RF^4}\frac{1}{k^2+m^2+i\epsilon}
\end{equation}
where we can notice the change $i\to -i$ 

 It is worth noting that no interaction can mix the components $\hat{\phi}_{+}$ and $\hat{\phi}_{-}$ because they are part of the direct sum formulation. For example,  a $\lambda\phi^3$ interaction is,  according to \eqref{disum}, 
\begin{equation}
	\Lc_{int} = -\frac{\lambda}{3}\hat{\phi^3} = -\frac{\lambda}{3} \begin{pmatrix}
		\hat{\phi}_{+}^3 & 0 \\ 
		0 & 	\hat{\phi}_{-}^3 
	\end{pmatrix}
\end{equation}
In DQFT, the standard model (SM) degrees of freedom, including particles $\vert SM\rangle$ and antiparticles $\vert \overline{SM}\rangle$ are written as direct-sum of two components corresponding to direct-sum vacuum:
\begin{equation}
    \vert 0_{SM}\rangle = \begin{pmatrix}
        \vert 0_{SM+}\rangle \\ 
        \vert 0_{SM-}\rangle 
    \end{pmatrix} \quad \vert SM\rangle = \frac{1}{\sqrt{2}}\begin{pmatrix}
        \vert SM_+\rangle \\ 
        \vert SM_-\rangle \end{pmatrix} \quad \vert \overline{SM}\rangle = \frac{1}{\sqrt{2}}\begin{pmatrix}
        \vert \overline{SM}_+\rangle \\ 
        \vert \overline{SM}_-\rangle 
    \end{pmatrix}
\end{equation}
We apply the same structure of the geometric superselection rule for all Fock spaces of the SM degrees of freedom. This would imply the definition of parity-conjugated regions for every state is the same. The~DQFT quantization of a real scalar field can be straightforwardly extended to complex scalars, fermions, and gauge fields. The framework here is every quantum field is represented as a direct-sum of two components, which are $\Pc\Tc$ mirror images of each other, covering the whole Minkowski spacetime. Therefore, the standard techniques of field quantization~\cite{Buchbinder:2021wzv} are adaptable to the DQFT. Below, we illustrate them in the context of complex scalar, fermion, and gauge fields. 
\subsubsection*{\textbf{Complex scalar field}:}

Complex scalar field operator $\hat{\phi}_c  = \frac{1}{\sqrt{2}}\LF \hat{\phi}_{c\,+}\oplus  \hat{\phi}_{c\,-}\RF$ in DQFT is expanded in terms of the four sets of creation and annihilation operators as 
\begin{equation}
	\begin{aligned}
		&  \hat{\phi}_{c\,\pm} = \int  \frac{d^3k}{\LF 2\pi \RF^{3/2}}\frac{1}{\sqrt{2\vert k_0\vert }}\Bigg[ a_{(\pm)\textbf{k}}e^{\pm ik\cdot  x}+b_{(\pm)\textbf{k}}^\dagger e^{\mp ik\cdot x}   \Bigg] \\ & \LT \hat{\phi}_{c\,+},\, \hat{\phi}_{c\,-} \RT =0\,,
	\end{aligned}
\end{equation}
{where $a_{(\pm)\textbf{k}},\, a^\dagger_{(\pm)\textbf{k}}$ and $b_{(\pm)\textbf{k}},\, b^\dagger_{(\pm)\textbf{k}}$ are canonical creation and annihilation operators of the parity conjugate regions (denoted by subscripts $_{(\pm)}$) attached with geometric superselection sector.
	All the cross commutation relations of \mbox{$a_{(\pm)},\, a^\dagger_{(\pm)}$ and $b_{(\pm)},\, b^\dagger_{(\pm)}$ vanish.}}

\subsubsection*{\textbf{Fermionic field}:}

Fermionic field operator $ \hat \psi = \frac{1}{\sqrt{2}}\LF \hat \psi_+\oplus \hat \psi_- \RF$ in DQFT is expanded as
\begin{equation}
	\hat  \psi_{\pm} = \sum_{{\tilde s}} \int \frac{d^3k}{\LF 2\pi \RF^{3/2}\sqrt{2\vert k_0\vert}} \Bigg[ c_{{\tilde s}(\pm)\textbf{k}} u_{\tilde s}(\textbf{k}) e^{\pm ik\cdot x} + d_{{\tilde s}(\pm)\textbf{k}}^\dagger v_{\tilde s}(\textbf{k}) e^{\mp ik\cdot x}\Bigg]
\end{equation}
where ${\tilde s}=1,2$ correspond to the two independent solutions of $\LF \slashed k+m\RF u_{\tilde s}=0$ and $\LF -\slashed k+m\RF v_{\tilde s}=0$ corresponding to spin-$\pm\frac{1}{2}$. The~creation and annihilation operators of the Fock space geometric superselection sector, satisfy the anti-commutation relations $\Big\{ c_{{\tilde s}(\pm)\textbf{k}},\,c_{{\tilde s}(\pm)\textbf{k}}^\dagger \Big\}=1,\, \Big\{ c_{{\tilde s}(\mp)\textbf{k}},\,c_{{\tilde s}(\pm)\textbf{k}}^\dagger \Big\}=\Big\{ c_{{\tilde s}(\mp)\textbf{k}},\,c_{{\tilde s}(\pm)\textbf{k}} \Big\}=0$ leading to the new causality condition $\Big\{ \hat\psi_+,\,\hat \psi_-\Big\} =0$.

\subsubsection*{\textbf{Gauge field}:}

The vector field operator $\hat A_\mu = \frac{1}{\sqrt{2}}\LF \hat{A}_{+\mu}\oplus \hat A_{-\mu} \RF$ in DQFT expressed as
\begin{equation}
	\hat{A}_{\pm \mu}= \int \frac{d^3k}{\LF 2\pi \RF^{3/2}\sqrt{2\vert k_0\vert }} e^{(\lambda)}_\mu\Bigg[ c_{(\pm \lambda)\textbf{k}} e^{\pm ik\cdot x}+c^\dagger_{(\pm \lambda)\textbf{k}} e^{\mp ik\cdot x}  \Bigg]
\end{equation} 
where $e^{(\lambda)}_\mu$ is the polarization vector satisfying the transverse and traceless conditions. The~creation and annihilation operators $c_{(\pm \lambda)\textbf{k}},\,c_{(\pm \lambda)\textbf{k}}^\dagger$ satisfy the similar relations \linebreak as \eqref{comcan}.

All of this demonstrates that all the standard model calculations remain unchanged, as the interaction terms are decomposed into a direct-sum structure in the following manner.
\begin{equation}
   \Lc_c \sim\Oc_{SM}^3=\begin{pmatrix}
        \Oc_{SM_+}^3 & 0 \\ 
        0 & \Oc_{SM_-}^3
    \end{pmatrix} \quad \Lc_q \sim \Oc_{SM}^4 = \begin{pmatrix}
        \Oc_{SM_+}^4 & 0 \\ 
        0 & \Oc_{SM_-}^4
    \end{pmatrix}
\end{equation}
Here, $\Oc_{SM}$ is an arbitrary operator involving any SM fields and their derivatives. {(}Remember that any derivative operators must be split into components joined by direct-sum operation{)}.  
The S-matrix also becomes the direct-sum with two parts
\begin{equation}
	\begin{aligned}
		S_T& = S_+ \oplus S_{-} 
	\end{aligned}
	\label{dQFTSM}
\end{equation}
where
\begin{equation}
	S_+= T_1 \Bigg\{  e^{-i\int_{-\infty}^{\infty} H_{int} \,\, dt } \Bigg \},\quad S_{-}  = T_2 \Bigg\{ e^{i\int_\infty^{-\infty} H_{int } \,\,  dt } \Bigg \}
	\label{DQFTSM12}
\end{equation}
with $T_1,\, T_2$ representing the time orderings attached to the respective Fock space arrow of time.
Therefore, we can deduce that the DQFT framework leaves QFT calculations in Minkowski spacetime unaffected, as~the spacetime itself is $\Pc\Tc$ symmetric. With~DQFT, any scattering amplitude, such as the transition from N particles to M particles, the~result remains the same as in standard QFT. Thus, the~amplitude of this in DQFT would become

\begin{equation}
	\begin{aligned}
    A_{N\to M} & = \frac{A^{N\to M}_+ \LF p_a, -p_b\RF  + A^{N\to M}_-\LF -p_a, p_b \RF}{2} \\  A^{N\to M}_+ \LF p_a, -p_b\RF & = A^{N\to M}_-\LF -p_a, p_b \RF,
    \end{aligned}
\end{equation}
where $p_a,\,p_b$ with $a=1,\cdots N$ and $b=1,\cdots M$ represent the 4-momenta of all the states involved in the scattering. 
$A_{\pm}$ represent amplitudes as a function of 4-momenta of initial and final states 
computed in both vacuums $\vert 0_{SM\pm}\rangle$. Notice that the in (out) states in $\vert 0_{SM\pm}\rangle$ come with the opposite sign, which is due to the arrow of time being opposite in both the vacuums. The~amplitudes $A_{\pm}$ are equal at any order in perturbation theory due to the $\Pc\Tc$ symmetry of Minkowski spacetime. 
The \cpt (charge conjugation, parity, and~time reversal) invariance of scattering amplitudes~\cite{Coleman:2018mew} valid in both vacuums, which means
\begin{equation}
    A^{N\to M}_+(p_a, -p_b) = A^{M\to N}_+(-p_a, p_b) , \quad A^{N\to M}_-(-p_a, p_b) = A^{M\to N}_-(p_a, -p_b) \,.
\end{equation}
This is because the \cpt operation on any scattering process would turn the outgoing anti-particles into in-going particles and vice-versa~\cite{Coleman:2018mew}. 

Aesthetically, in~DQFT, we have N particles (quantum fields) going from $t_p\to -\infty$, according to Fock space $\Fc_{+}$, while according to Fock space $\Fc_{-}$ those N particles are going from $t_p\to +\infty$. After~scattering we have outgoing $M$ particles at  $t_p\to +\infty$ with respect to $\Fc_{+}$ while those $M$ particles appear at $t_p\to -\infty$ according to $\Fc_{-}$. Now, one may wonder what the (imaginary) observer's arrow of time is or in what direction (of time) the observer sees the scattering. Is it $t_p: -\infty \to \infty$ or $t_p: +\infty \to -\infty$?  First, we remind the reader that an observer cannot be defined in quantum theory because {time} is a parameter (and spatial position is an operator). Thus, time has a special meaning in quantum theory, as~we discussed earlier. Therefore, one should distinguish the observer's clock from the quantum mechanical notion of {time}. Of~course, in~a particle physics laboratory (for example, the~Large Hadron Collider), we are the observers seeing particle scatterings; however, what we measure is only pre- and post-scattering states and the corresponding cross-sections. Thus, one need not worry about any notion of (quantum mechanical) time there. Here, the~direction in which the scattering proceeds is just $N$ particles interacting and scattering into $M$ particles. So, we have initial and final states that define the scattering process direction in what we can call the observer's (classical) arrow of~time.   
  
In summary, we introduced a novel approach to quantum field theory by establishing a direct-sum mathematical framework that bridges $\Pc\Tc$-conjugate sheets of spacetime. Through the use of geometric superselection rules defined by parity-conjugate regions of physical space, we integrated the concept of two distinct arrows of time within a unified quantum state description. While DQFT preserves the practical outcomes of standard model particle physics, it offers a deeper insight into the role of ``time'' in quantum theory, providing a fresh perspective on its fundamental nature.
This new understanding set up a stage for the consistent QFT in curved spacetime, as we will explore in the later sections.

\subsection{Conformal diagram for DQFT in Minkowski spacetime}
\label{sec:NPdiagrams}
 
In this section, the idea is to give a compact diagrammatic description of our DQFT.   Since this pictorial representation is not about classical theory, we want to represent  QFT, discrete symmetries, and the direct-sum vacuum in the {conformal diagrams}.

Minkowski space-time \eqref{min} is written in terms of the compactified coordinates by \cite{Griffiths:2009dfa}
\begin{equation}
	ds^2 = \frac{4}{\xi(T_p,R)}\LF-dT_p^2+ dR^2 + \sin^2(R) d\Omega^2 \RF\,, 
	\label{Minkowski-TR}
\end{equation}
where $-\pi+R <T_p<\pi-R$ and $0\leq R<\pi$, and $\xi(T_p,R) = \LF \cos T_p+\cos R\RF^2 $ is a positive function and $d\Omega^2$ represents the line element of two sphere. 
\begin{equation}
	\begin{aligned}
		T_p  & = \arctan\LF t_p+r \RF + \arctan\LF t_p-r \RF \\
		R & = 	\arctan\LF t_p+r \RF -\arctan\LF t_p-r \RF 
	\end{aligned}
	\label{TRplane}
\end{equation}
We can notice that under time reflection operation $\Tc$, we get
\begin{equation}
	\Tc: t_p\to -t_p \implies T_p\to -T_p, \quad R\to R\,.
\end{equation}
\begin{figure}[ht!]
	\centering
	\includegraphics[width=0.7\linewidth]{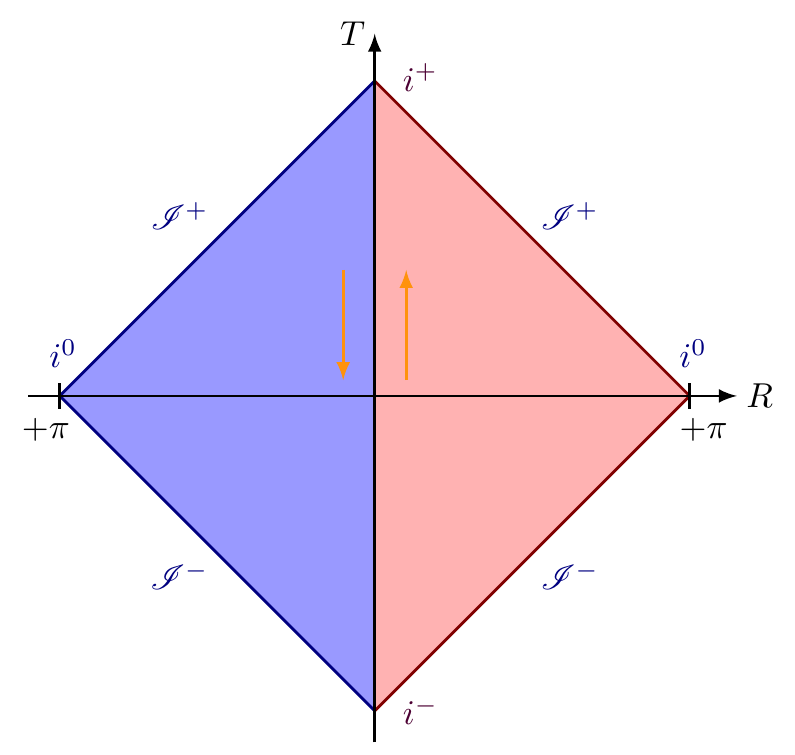}
	\caption{
	Minkowski spacetime conformal diagram that represents the DQFT construction with direct-sum vacuum \eqref{tFs} as discussed in Sec.~\ref{sec:direct-sumFQFT}. The left and right triangles are $\Pc\Tc$ conjugates of each other. The DQFT quantum field operator  $\hat{\phi}$ is the direct-sum of $\hat{\phi}_+, \hat{\phi}_{-}$ defined on the right and left, respectively, which follows from \eqref{disum}. Each point in the blue triangle is the parity conjugate of the one in the red triangle. The arrows in the figure indicate the arrows of time $t_p: \mp \infty \to \pm \infty $.  }
	\label{fig:minkowski-np}
\end{figure}
We now use the compactified coordinates  \eqref{TRplane} to draw a conformal diagram that represents DQFT construction \eqref{tFs}. As depicted in Fig.~\ref{fig:minkowski-np}, in DQFT construction, we split space into two parts by parity $\Pc$ transformation. Therefore, the right triangles in red and blue are parity conjugate regions. The quantum states, in the whole spacetime, are a direct-sum of states $\hat{\phi}_+$, evolving with respect to an arrow of time $t_p: -\infty \to \infty $ (in the red triangle), and states $\hat{\phi}_{-}$ evolving backward in time i.e., $t_p: \infty \to -\infty$ (in the blue triangle). As a consequence of \eqref{comcan},  we have
\begin{equation}
\left[	\Oc_+\LF t_p,\,\textbf{x} \RF,\, \Oc_{-}\LF -t_p^\prime,\, -\textbf{x}^\prime \RF\right] = 0\,. 
\label{Minloc}
\end{equation} 
Since we also demand
\begin{equation}
\begin{aligned}
	\left[\Oc_+\LF x \RF,\, \Oc_{+}(y)\right] & =0,\quad \LF x-y \RF^2>0\, \\
		\left[\Oc_{-}\LF -x \RF,\, \Oc_{-}(-y)\right] & =0,\quad \LF x-y \RF^2>0\,.
	\end{aligned}
\label{causalityeq}
\end{equation} 
Therefore, as a consequence of \eqref{Minloc} and \eqref{causalityeq}, we satisfy both locality and causality requirements in DQFT. 


\section{De Sitter spacetime and quantization}

\label{sec6DQFTDS}

After Minkowski spacetime, the physically most relevant maximally symmetric spacetime is dS. The dS metric exhibits the presence of event or apparent horizons, depending on the coordinate system we choose to represent it. As such, it is most often said that the (quantum) physics in dS spacetime is observer dependent \cite{Hartong:2004rra}.  
There are two main questions involved here. The first question is: what are the rules of quantization in dS spacetime, which should be addressed within the validity of perturbation theory? And this question actually seeks to understand how to combine gravity and quantum mechanics. The second question is: how can one end up with a dS spacetime starting from the grand framework of quantum gravity? Which has been the subject of investigation within the scope of string theory for several decades \cite{Witten:2001kn}. Our efforts are directed to find an answer to, or at least a consistent route to answer, the first question, which we think will ultimately be helpful to attack the second grand question. Surely, we greatly acknowledge the weight of the second question, but we stress that it is beyond the scope of this paper. 

Without further delay, let us turn to the main subject of this section, where our aim is to apply a set of new quantization rules to de~Sitter (dS) spacetime.
Those will play a vital role in our understanding of cosmology. The success of inflationary cosmology is a consequence of the spontaneous breaking of de~Sitter symmetry \cite{Starobinsky:1980te}. The QFT in dS construction, considered here, is different from the well-known ones in the literature \cite{Birrell:1982ix,Mukhanov:2007zz} which lead to inconsistencies in relation to unitarity (evolution of pure states into mixed states) and information loss \cite{Parikh_2003}. Another important issue of dS SQFT is that one can only define {\it in} or {\it out} states, which forbids us to define a consistent scattering problem. Nevertheless, several developments were proposed following an unknown theory of quantum gravity, or based on entanglement between unknown Universes, despite admitting the lack of an S-matrix formulation in dS space \cite{Witten:2001kn,Bousso:2004tv,Balasubramanian:2020xqf,Balasubramanian:2021wgd,Anninos:2012qw,Aalsma:2021kle,Goheer:2002vf,Spradlin:2001pw,Hartman:2020khs}. 
The story of (quantum) troubles with dS spacetime \cite{Goheer:2002vf} is not really an issue raised in recent years, but rather started from the days of Schr\"{o}dinger. Indeed, he already pointed out that dS spacetime has ambiguities, which become a serious obstacle to constructing a reasonable quantum theory \cite{Schrodinger1956}. 
Schr\"{o}dinger's monograph named \ says {Expanding Universes} \cite{Schrodinger1956} points to the non-existence of a global arrow of time or, in other words, the problem of the lack of a Cauchy surface in dS spacetime, which hampers a correct quantum theory formulation. Schr\"{o}dinger proposed that one must carry with the so-called antipodal identification of the points in spacetime to define one Universe with a unique arrow of time and a well-defined Cauchy surface.  This Schr\"{o}dinger view is known as Elliptic dS, and it has been explored to define a dS unitary QFT \cite{SANCHEZ19871111,Parikh_2003}. However, the idea of Elliptic dS space fails to give a consistent QFTCS due to the possibility of closed time-like curve occurrence if one takes a departure from dS spacetime \cite{Aguirre_2003}. Nevertheless, Sch\"{o}dinger monograph provides inspiring ideas, such as distinguishing the notions of time classically and quantum mechanically, which we strongly envisage as the key for formulating QFTCS. Furthermore, Schr\"{o}dinger proposes that any observer should have complete information within his/her own horizon, which in modern language translates to the description of the Universe within the horizon using so-called pure states. We take this as a guiding principle in the present work.

The question we address, in the context of dS spacetime in a given coordinate system, is how can we quantize a free KG field, and how do we use the understanding of free fields in that spacetime to define a notion of scattering if we include the interactions perturbatively. However, we surely ignore here the quantum fields backreaction on the dS geometry and also leave the questions of whether dS is stable or not, and if one does require to understand (quantum) gravity at non-perturbative level (See \cite{Akhmedov:2013vka} for more discussion on these topics).  We strongly acknowledge the later questions are legitimate, but we strictly limit ourselves to QFTCS rather than quantum gravity. 
Below, we present our analysis for dS QFT, expressed in FLRW coordinates. Part of what we will discuss can also be found in \cite{Kumar:2022zff,Gaztanaga:2024vtr,GKM}. Therein, we extract observational consequences of applying QFTCS with direct-sum Fock space in the context of single-field inflationary cosmology addressing one of the prominent CMB anomalies. 
In other words, direct-sum QFTCS allows us to make a prediction that can be tested with future CMB and Primordial Gravitational Waves (PGWs) observations \cite{Kumar:2022zff}. This paper aims to provide further theoretical ground and take a step towards understanding unitarity and the S-matrix in QFTCS. 

\section{Direct-sum QFT in de Sitter spacetime}

\label{sec:dQFTinDs}

In this section, we first address the subject DQFT with KG field quantization in flat FLRW dS spacetime. Our approach applies a similar procedure to the DQFT framework we discussed in Sections \ref{sec:direct-sumFQFT} and \ref{sec:NPdiagrams} in the context of Minkowski spacetime. At this point, we consider DQFT quantization in flat (expanding) FLRW dS spacetime \eqref{expcon} and a similar procedure can be followed, using discrete spacetime transformations, for closed dS according to \eqref{discflrw} and \eqref{clldS}. 

The action of a massless scalar field in dS (flat FLRW case) is given by
\begin{equation}
	S_{\phi} = -\frac{1}{2}\int d\tau d^3x a^2	\phi\LF \pd_\tau^2+2\Hc \pd_\tau -\pd_i^2 \RF \phi\,, 
\end{equation}
where $\Hc = \frac{1}{a}\frac{da}{d\tau}=aH$. 
We can verify that the above action is invariant under the $\Pc\Tc$ transformation, defined in \eqref{disflat}. 
To implement quantization, it is convenient to rescale the field $\phi\to a \phi$,  which allows to define the following action 
\begin{equation}
	S_\phi = \frac{1}{2} \int d\tau d^3x \Big[ \phi^{\prime 2} - \LF \pd_i\phi \RF^2  +  \frac{2}{\tau^2}  \phi^2 \Big]\,. 
	\label{msaction}
\end{equation}
The above action describes a KG field in flat space-time with a time-dependent (negative) mass $\mu^2 =  -\frac{2}{\tau^2}<0$. The physics of quantum fields in dS is also interestingly related to the quantum aspects of inverted harmonic oscillators which is explicitly unveiled in \cite{GKM} in the framework of direct-sum quantum theory. 

When we quantize this scalar field (by following a procedure employed in Sec.~\ref{sec:direct-sumFQFT}), the scalar field operator $\hat{\phi}\LF \tau,\, \textbf{x} \RF$  is written as a direct-sum of two components as
\begin{equation}
	\hat{\phi}\LF \tau,\, \textbf{x} \RF =  \frac{1}{\sqrt{2}}	\hat{\phi}_I\LF \tau,\, \textbf{x} \RF \oplus 	\frac{1}{\sqrt{2}}\hat{\phi}_{II}\LF -\tau,\, -\textbf{x} \RF = \frac{1}{\sqrt{2}}\begin{pmatrix}
		\hat{\phi}_I\LF \tau,\,\textbf{x} \RF & 0 \\ 0 & \hat{\phi}_{II}\LF -\tau,\,-\textbf{x} \RF
	\end{pmatrix} \,. 
\label{disumdS}
\end{equation}
where 
\begin{equation}
	\begin{aligned}
			\hat{\phi}_I\LF \tau,\, \textbf{x} \RF &   = \frac{1}{\LF 2\pi \RF^{3/2}}\int  d^3k\Bigg[ \hat{c}_{I\,\textbf{k}} {\phi}_{I\,k}\LF \tau \RF e^{-i\textbf{k}\cdot \textbf{x}} + \hat{c}_{I\,\textbf{k}}^\dagger {\phi}^\ast_{I\,k}\LF \tau \RF e^{i\textbf{k}\cdot \textbf{x}} \Bigg]\, \\ 
						\hat{\phi}_{II}\LF -\tau,\, -\textbf{x} \RF &   = \frac{1}{\LF 2\pi \RF^{3/2}}\int   d^3k\Bigg[ \hat{c}_{II\,\textbf{k}} {\phi}_{-\,k}\LF -\tau \RF e^{i\textbf{k}\cdot \textbf{x}} + \hat{c}_{II\,\textbf{k}}^\dagger {\phi}^\ast_{II\,k}\LF -\tau \RF e^{-i\textbf{k}\cdot \textbf{x}} \Bigg]\, 
	\end{aligned}
	\label{vdfieldS}
\end{equation}
describe the quantum field in the parity conjugate regions of Minkowski space with opposite arrows of time ($\tau: -\infty \to 0$ and $\tau: \infty \to 0$) in the geometric superselection sector Fock spaces 
\begin{equation}
    \Fc  = \Fc_{dSI}\oplus \Fc_{dSII}\,,
    \label{dSFock}
\end{equation}
where the creation and annihilation operators $\hat{c}_{\pm\textbf{k}},\, \hat{c}^\dagger_{\pm\textbf{k}}$ satisfy the following commutation relations (which are very similar to \eqref{comcan})
\begin{equation}
	\begin{aligned}
		[\hat{c}_{I\,\textbf{k}},\,\hat{c}_{I\,\textbf{k}^\prime}^\dagger] & = 	[\hat{c}_{II\,\textbf{k}},\,\hat{c}_{II\,\textbf{k}^\prime}^\dagger] = \delta^{(3)}\LF \textbf{k}-\textbf{k}^\prime \RF\\
		[\hat{c}_{I\,\textbf{k}},\,\hat{c}_{II\,\textbf{k}^\prime}] &=	[\hat{c}_{I\,\textbf{k}},\,\hat{c}_{II\,\textbf{k}^\prime}^\dagger] = [\hat{c}_{I\,\textbf{k}}^\dagger,\,\hat{c}_{II\,\textbf{k}^\prime}^\dagger]  =0\,.
	\end{aligned}
	\label{comcandS}
\end{equation}
Consequently, we also have
\begin{equation}
	\begin{aligned}
		[\hat{	\phi}_I\LF \tau, \textbf{x} \RF,\,\pi_I\LF \tau,\,\textbf{x}^\prime \RF] &  = i \delta\LF \textbf{x}-\textbf{x}^\prime \RF\\
		 [\hat{	\phi}_{II}\LF- \tau, -\textbf{x} \RF,\,\pi_{II}\LF -\tau,\,-\textbf{x}^\prime \RF] & = -i \delta\LF \textbf{x}-\textbf{x}^\prime \RF \\
	\LT \hat{\phi}_I\LF \tau,\,\textbf{x} \RF,\,\hat{\phi}_{II}\LF -\tau,\,-\textbf{x} \RF  \RT & = 0\,,
	\label{canDQFTds}
	\end{aligned}
\end{equation}
where 
\begin{equation}
	\pi_I \LF \tau,\,\textbf{x} \RF = \pd_\tau \phi_I,\quad  	\pi_{II} \LF -\tau,\,-\textbf{x} \RF = -\pd_\tau\phi_{II}\,.  
\end{equation}
The third line of \eqref{canDQFTds} is consistent with the quantum theory locality and causality principle. {It is true for any operator in Region I and Region II of dS spacetime
\begin{equation}
\left[	\Oc_I\LF \tau,\,\textbf{x} \RF,\, \Oc_{II}\LF -\tau^\prime,\, -\textbf{x}^\prime \RF\right] = 0\,. 
\label{localitydS}
\end{equation}
It is easy to see \eqref{localitydS} is analogous to the one of DQFT in Minkowski spacetime \eqref{Minloc}. }

In flat FLRW dS, as the Universe expands, modes exit the horizon at spacelike separated points. According to our description, the mode exiting the horizon at a point $\LF \theta,\,\varphi \RF$ fixes the state exiting on the parity conjugate point  $\LF \pi-\theta,\,\pi+\varphi \RF$ without violating the locality and causality principle guaranteed by the commutation relations. We shall discuss this further in the later section. 

Functions $\phi_{I\,k},\,\phi_{II\,k}$ are the mode functions 
\begin{equation}
	\begin{aligned}
	\phi_{I,\,k}  & = \alpha_{I\,k} \frac{e^{-ik\tau}}{\sqrt{2k}}\LF 1-\frac{i}{k\tau} \RF +\beta_{I\,k} \frac{e^{ik\tau}}{\sqrt{2k}}\LF 1+\frac{i}{k\tau} \RF\,, \\
		\phi_{II,\,k}  & = \alpha_{II\,k} \frac{e^{ik\tau}}{\sqrt{2k}}\LF 1+\frac{i}{k\tau} \RF +\beta_{II\,k} \frac{e^{-ik\tau}}{\sqrt{2k}}\LF 1-\frac{i}{k\tau} \RF\,, 
	\label{f1}
	\end{aligned}
\end{equation}
satisfying $\phi_{I,\,k} \Big\vert_{\Tc:\,\tau \to -\tau} = \phi_{II,\,k}$ and are obtained by solving the Mukhanov-Sasaki equation \cite{Kumar:2022zff}
\begin{equation}
	\begin{aligned}
	\phi_{m\,k}^{\prime\prime}+ \LF k^2-\frac{2}{\tau^2} \RF \phi_{m\,k}=0\,. 
	\end{aligned}
\label{MSeq}
\end{equation}
where $m= I,\,II$. We can notice, from \eqref{MSeq}, that in the limit $k^2\gg 2/\tau^2$ or $k^2\gg 2/r_H^2$, i.e., when the comoving wavelength of the mode is much less than the size of the comoving horizon $r_H$, we recover DQFT results discussed in Sec.~\ref{sec:direct-sumFQFT}. 
Therefore, we fix the Bogoliubov coefficients as $\LF \alpha_{I\,k},\, \beta_{I\,k}\RF = \LF \alpha_{II\,k},\, \beta_{II\,k}\RF  = \LF 1,\,0 \RF$, which are compatible with the Wronskian conditions $	\phi_{I,\,k} 	\phi^{\prime\ast}_{I,\,k}-	\phi^\ast_{I,\,k}	\varphi^\prime_{I,\,k} = i $ and $	\phi_{II,\,k} 	\phi^{\prime\ast}_{II,\,k}-	\phi^\ast_{II,\,k}	\phi^\prime_{II,\,k} = -i $, that corresponds to the canonical commutation relations \eqref{canDQFTds}. 

The dS spacetime vacuum is given by the direct sum
\begin{equation}
	\vert 0\rangle_{dS} = \vert 0\rangle_{dSI} \oplus \vert 0 \rangle_{dSII} = \begin{pmatrix}
		\vert 0 \rangle_{dSI} \\ 
		\vert 0 \rangle_{dSII}
	\end{pmatrix}\,.
\label{dsumFds}
\end{equation}
where $\vert 0\rangle_{dSI},\,\vert 0 \rangle_{dSII}$ are defined according to 
\begin{equation}
\hat{c}_{I\,\textbf{k}}\vert 0 \rangle_{dSI} = 0,\quad \hat{c}_{II\,\textbf{k}}\vert 0 \rangle_{dSII} = 0
\end{equation}
The choice $\LF \alpha_{I\,k},\, \beta_{I\,k}\RF = \LF 1,\,0 \RF = \LF \alpha_{II\,k},\, \beta_{II\,k}\RF$ represents direct-sum analog of Bunch-Davies vacuum.  
Because, the limits $\tau \to \pm \infty$ which mean $k^2\gg 2/\tau^2$ are nothing arriving to the local Minkowski limit. Therefore, in these limits, dS vacuum \eqref{dsumFds} in DQFT should exactly reduce to the Minkowski vacuum \eqref{tFs}. To the field operators, this would mean
\begin{equation}
  \hat  \phi_{dSI}\Big\vert_{\tau\to -\infty} \equiv \hat \phi_{+},\quad \hat  \phi_{dSII}\Big\vert_{\tau\to +\infty} \equiv \hat \phi_{-}
\end{equation}
Notice that, in the DQFT framework, we divide the spatial part of dS expanding Universe \eqref{expcon} into two parts, $dSI$ and $dSII$, through the parity transformation, and then we associate opposite time evolutions to quantum states in region $dSI$  ($\tau: -\infty \to 0$) and  $dSII$ ($\tau: \infty \to 0$). 
The direct-sum of these component states represents the full (quantum mechanical) evolution of the scalar field in the direct-sum vacuum \eqref{dsumFds}, which describes the whole dS spacetime (expanding Universe). 
In other words, the vacuum \eqref{dsumFds} satisfy the discrete spacetime symmetry \eqref{dSsym}.
We argue that only a single degree of freedom is represented as the direct-sum of two complementary field components $\hat \phi_{I},\,\hat \phi_{II}$. We can verify this by observing that 
\begin{equation}
	\begin{aligned}
		& \frac{k^3}{a^2}{}_{dSI}\langle 0\vert \hat{\phi}_{I}\LF \tau,\, \textbf{x} \RF \hat{\phi}_{I}\LF \tau,\, \textbf{x}^\prime \RF\vert 0\rangle_{dSI} = \\ & \frac{k^3}{a^2} {}_{dSII}\langle 0\vert \hat{\phi}_{II}\LF -\tau,\, -\textbf{x} \RF \hat{\phi}_{II}\LF -\tau,\, -\textbf{x}^\prime \RF\vert 0\rangle_{dSII} = \frac{H^2}{4\pi^2}\,. 
	\end{aligned}
	\label{eqcorrdS}
\end{equation}
in the limit $k\vert\tau\vert  \to 0 $.

We see that \eqref{eqcorrdS}, analogous to the result of  Minkowski spacetime DQFT \eqref{CPTnewdef}, implies that the correlations of quantum fields in the parity conjugate regions are identical in the case of dS spacetime.  It was recently shown that the breakdown on the equality in \eqref{eqcorrdS} for quantum fields during the inflationary (quasi-dS) expansion leads to parity asymmetry in the cosmic microwave background \cite{Gaztanaga:2024vtr}.

In the later section, we conceptually deal with understanding the scattering problem and how dS DQFT enables us to define an S-matrix, which is thought to be an impossible task to achieve for dS spacetime \cite{Bousso:2004tv}. 

\subsection{Unitarity, observer complementarity  in de Sitter DQFT}

One of the issues with dS spacetime is: it challenges us with unitarity violation, for example, with the evolution of pure states into mixed states. In addition, the thermodynamic point of view that emerges with the finiteness of entropy, which leads to the suggestion of Hilbert space finiteness \cite{Gibbons:1977mu,Parikh_2005,Dyson:2002nt}. All these troubling issues emerge at the fundamental level because we do not know how to reconstruct what is beyond the observer's horizon. One of the resolutions to the problem is the proposal of dS spacetime observer complementarity \cite{Dyson:2002nt},\footnote{It is motivated by BH complementarity principle \cite{Susskind:1993if,tHooft:1984kcu} which we studied in a separate \cite{Kumar:2023hbj}. }, which states that all observers are equivalent (no one should see any violation of unitarity), and all the information is contained within each observer horizon. This is known as the central dogma in dS, which is mainly explored in the context of static dS in string theory and AdS/CFT \cite{Shaghoulian:2021cef}. We aim to address the above issues from the point of view of dS DQFT. 
\begin{figure}
	\centering
\includegraphics[width=0.7\linewidth]{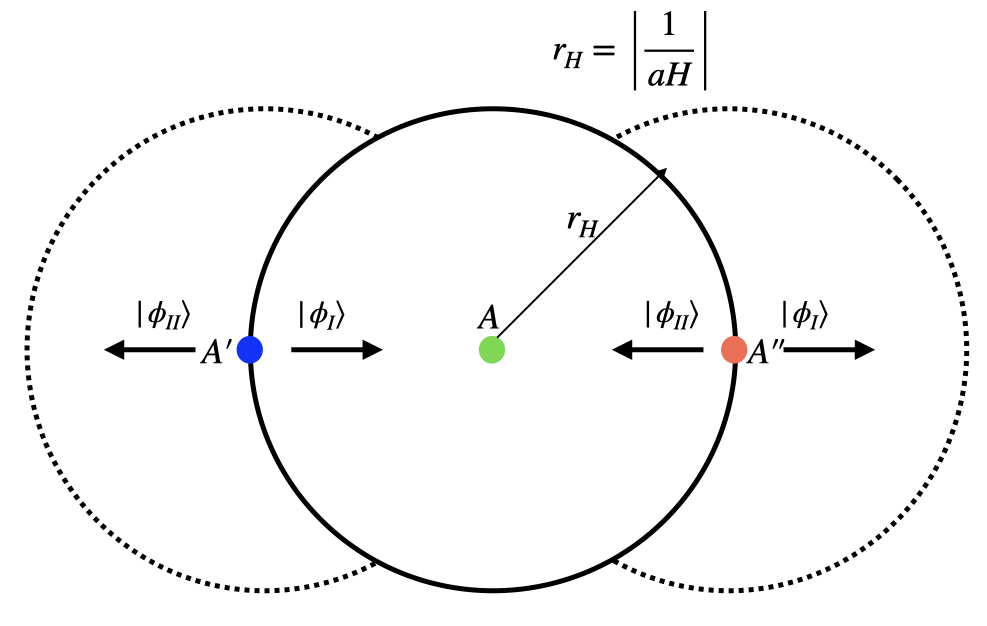}
	\caption{This figure illustrates three comoving (imaginary) observers $A,\,A^\prime,\, A^{\prime\prime}$ with the corresponding comoving horizon radius $r_H= \vert \frac{1}{aH}\vert$ at a moment of dS expansion. We have suppressed here the angular coordinates $\LF \theta,\,\varphi \RF$. Points $A^\prime$ and $A^{\prime\prime}$ are on the horizon of $A$ at antipodal sides of the horizon, that are spacelike separated, i.e., at the angles $\LF \theta,\,\varphi \RF$ and $\LF \pi-\theta,\, \pi+\varphi \RF$ respectively. The dotted circles with the same $r_H$ represent the respective comoving horizons of $A^\prime$ and $A^{\prime\prime}$. In the figure, $\vert \phi_I\rangle ,\,\vert \phi_{II}\rangle $ are defined in \eqref{dSdiaSt}.  }
	\label{fig:desitteru}
\end{figure}

First, the comoving horizon \eqref{horizonds} is the most important quantity attached to dS spacetime. We represent it in Fig.~\ref{fig:desitteru}, where we take an (imaginary) observer $A$ with $r_H$ at a moment of dS expansion, and two other imaginary observers $A^\prime,\, A^{\prime\prime}$ on the horizon of $A$. Expanding the dS Universe would translate to shrinking the comoving horizon size as the scale factor grows according to the thermodynamic concept of time.  In the framework of dS DQFT, we divide space into two halves, related by parity transformation (a discrete transformation and in 3D it cannot be related to any continues rotations), and associate (quantum mechanical) time evolution to the direct-sum state in the direct-sum Fock space \eqref{dsumFds}. This represents a single degree of freedom (following from \eqref{disumdS}) in the entire two spatial regions, which span points of $\LF \theta,\, \varphi \RF$ and $\LF \pi-\theta,\, \pi+\varphi \RF$ within the comoving radius $r_H$.  In Fig.~\ref{fig:desitteru}, $\vert \phi_I \rangle = \hat{c}_{I\,\textbf{k}}^\dagger \vert 0\rangle_{dSI}$ and $\vert \phi_{II} \rangle = \hat{c}_{II\,\textbf{k}}^\dagger \vert 0\rangle_{dSII}$ form the direct-sum state
\begin{equation}
	\begin{aligned}
		\vert \phi \rangle & = \LF \hat{c}^\dagger_{I\,\textbf{k}} \oplus \hat{c}^\dagger_{II\,\textbf{k}} \RF \vert 0\rangle_{dS} \\ & = \frac{1}{\sqrt{2}}\begin{pmatrix}
			 \hat{c}^\dagger_{I\,\textbf{k}} & 0 \\ 
			 0 &  \hat{c}^\dagger_{II\,\textbf{k}} 
		\end{pmatrix}  \begin{pmatrix}
		\vert 0\rangle_{dSI} \\
		\vert 0\rangle_{dSII}
	\end{pmatrix} \\ & 
		= \frac{1}{\sqrt{2}} \vert \phi_I\rangle \oplus \frac{1}{\sqrt{2}} \vert \phi_{II}\rangle = \frac{1}{\sqrt{2}}\begin{pmatrix}
			\vert \phi_I\rangle \\ 
			\vert \phi_{II} \rangle
		\end{pmatrix}\,. 
	\end{aligned}
\label{dSdiaSt}
\end{equation} 
By construction, in dS  DQFT, which we established in Sec.~\ref{sec:dQFTinDs} and \eqref{disumdS}, the imaginary observer $A$ always witnesses the direct-sum state \eqref{dSdiaSt} within his (comoving) horizon and, as we can see from Fig.~\ref{fig:desitteru}, the two imaginary observers at $A^\prime$ and $A^{\prime\prime}$ also have the same state within their horizon radius $r_H$. This implies that, by observing the direct-sum state \eqref{dSdiaSt}, the imaginary observer $A$ can reconstruct what is happening beyond his/her horizon.
Thus, we account for the observer complementarity principle, and every observer experiences the same physics in the sense that there will not be any evolution of pure states to mixed states, which is the problem according to the standard description of QFTCS.

{ In Fig.~\ref{fig:desitteru} we can see that, even though the imaginary observers $A^\prime$ and $A^{\prime\prime}$ are spacelike separated, $A^{\prime\prime}$ by finding the state $\vert \phi_{I}\rangle$ in the right hand side can deduce what $A^\prime$ would find on the left hand side. This implies $A^{\prime\prime}$ can accurately deduce what $A^\prime$ can access. This has profound implications for the understanding of dS spacetime. It has been recently pointed out that the deepest problem in understanding quantum \& gravity is the violation of unitarity at distances much larger than Planck scales, and it is also argued that the challenge for QFTCS is the conflict between quantum mechanics, relativity, and locality. To resolve the conundrums, it has been speculated that the {nonlocality} via quantum entanglement and direct-product Hilbert space would play a crucial role in formulating QFTCS \cite{Giddings:2022jda}. Here, we argue that a unitary QFTCS only requires a new concept of (local) time, which is a parameter, and an additional condition to the usual definition of {locality} is required (See \eqref{Minloc} and \eqref{localitydS}). Our QFA approach is based on the fact that there is no global definition of time in curved spacetime. Therefore, we completely rely on the discrete spacetime transformations; whenever one goes from one spatial region to another by parity $\textbf{x}\to -\textbf{x}$, we change from one geometric superselection to another and apply time reversal to the quantum states. The two Fock spaces direct-sum form the full description of a quantum state within the observable space associated with an imaginary observer.  Indeed,  according to \eqref{dSdiaSt}, a quantum state $\vert \phi\rangle$ means, $\vert \phi_I\rangle$ at position $\textbf{x}$ evolves forward in time (i.e., $\tau: -\infty\to 0$) and $\vert \phi_{II}\rangle$ at position $-\textbf{x}$ evolves backward in time (i.e., $\tau: \infty\to 0$) like they are two sides of the same coin. This can be generalized to multiparticle states as well following \eqref{disumFock} and \eqref{dsumFds}. For example, let us consider the usual two-particle states with the tensor product of Hilbert spaces; in the direct-sum formulation, it can be expressed as 
	\begin{equation}
		\begin{aligned}
		\Hc_P & = \Hc_{dS1} \otimes \Hc_{dS2} \\ 
		& = \LF \Hc_{dSI1}\oplus \Hc_{dSII1}\RF  \otimes  \LF \Hc_{dSI2}\oplus \Hc_{dSII2}\RF \\ 
		& = \LF \Hc_{dSI1}\otimes \Hc_{dSI2}\RF  \oplus  \LF \Hc_{dSII1}\otimes \Hc_{dSII2} \RF\,, 
		\end{aligned}
	\label{tenprod-Hilbert}
	\end{equation}
where we can see the two-particle tensor product Hilbert space is $\Pc\Tc$ symmetric.
Suppose we consider 
two-particle maximally entangled state $\in \Hc_P$ in the spacetime bounded by $r_H$,
\begin{equation}
    \vert \psi_{12}\rangle = \sum_{m,n} d_{mn} \vert \phi_{1} \rangle \otimes  \vert \phi_{2}\rangle \,,
\end{equation}
where $d_{mn} \neq d_m d_n$, $\vert \phi_1 \rangle = \sum_m d_m\vert \phi_{m1}\rangle $ and $\vert \phi_2 \rangle = \sum_n d_n\vert \phi_{n2}\rangle $.

According to DQFT, any state must have two components, which means 
\begin{equation}
    \vert \psi_{12}\rangle = \frac{1}{\sqrt{2}}\begin{pmatrix}
        \vert \psi_{I(12)}\rangle \\ 
        \vert \psi_{II(12)}\rangle 
    \end{pmatrix} = \frac{1}{\sqrt{2}}\begin{pmatrix}
        \sum_{m,n} d_{m,n} \vert \phi_{I1} \rangle \otimes  \vert \phi_{I2}\rangle \\ 
        \sum_{m,n} d_{m,n} \vert \phi_{II\,1} \rangle \otimes  \vert \phi_{II\,2}\rangle
    \end{pmatrix} 
    \label{puredS}
\end{equation}
{The state $ \vert \psi_{12}\rangle $ is a pure state because its density matrix is direct-sum of two pure states in the sectorial Hilbert spaces 
\begin{equation}
    \rho_{12} = \frac{1}{2} \rho_{I(12)}\oplus \frac{1}{2} \rho_{II(12)} 
\end{equation}
and their Von Neumann entropies vanish 
\begin{equation}
    S_{NI} = -Tr\LF \rho_{I(12)}\ln \rho_{I(12)}\RF=0,\quad S_{NII} = -Tr\LF \rho_{II(12)}\ln \rho_{II(12)}\RF=0\,.
\end{equation}}

This entangled pure state 
splits into a direct-sum of two pure state components of Hilbert spaces ($\Hc_{dSI}$ and $\Hc_{dSII}$) in the region $dSI$ and region $dSII$, respectively. 
This DQFT prescription by construction fully encodes the (quantum) information beyond the horizon to within the horizon so that an observer always witnesses pure states evolving into pure states within his/her horizon. This is solely due to the DQFT construction where an additional locality and causality condition \eqref{localitydS} is implemented. 
	  As we can see in Fig.~\ref{fig:desitteru}, the imaginary observer $A$ sees the combination of both $\vert \phi_I\rangle$ and $\vert \phi_{II}\rangle$ forming the full direct-sum state $\vert \phi\rangle$. On the right-hand side of $A$, even though $\vert \phi\rangle_I$ is going out of the horizon, the imaginary observer $A$ is not losing anything as he/she got to receive it from the left-hand side. At first sight, it may appear to be a duplication of information/cloning. But, it is absolutely not because no observer can see the same information twice within his/her observable horizon. Therefore, we definitely satisfy the so-called no-cloning theorem in quantum theory. 
	This imply all the imaginary observers in dS spacetime in the context of DQFT are equivalent although not identical. }
 
\subsection{DQFT in Static de Sitter space}

In this section, we qualitatively implement the scheme of dS DQFT in the static dS space represented by \eqref{statdS}. The Kruskal coordinates $\LF \Tilde{U},\,\Tilde{V} \RF$ are suitable for quantizing fields in curved spacetime because of coordinate singularity present in the coordinates $\LF t_s,\,r,\,\theta\, \varphi \RF$  \cite{Gibbons:1977mu,Kumar:2024oxf}. 
The dS horizon in these coordinates is traded with discrete transformations on  $\LF \Tilde{U},\,\Tilde{V} \RF$ \cite{Griffiths:2009dfa,Hartman:2017,Spradlin:2001pw} 
\begin{equation}
    \begin{aligned}
       \Uc  &= -e^{-H \bar{u}}<0,\quad &&\Vc= e^{H \bar{v}}>0\quad &&(\rm Region\,I)\\
        \Uc &= e^{-H \bar{u}}>0,\quad &&\Vc= -e^{H \bar{v}}<0\quad &&(\rm Region\, II) \\
         \Uc &= e^{-H \bar{u}}>0,\quad &&\Vc= e^{H \bar{v}}>0\quad &&(\rm Region\, III) \\
          \Uc &= -e^{-H \bar{u}}<0,\quad &&\Vc= -e^{H \bar{v}}<0\quad &&(\rm Region\, IV)
    \end{aligned}
    \label{UVKruskaldS}
\end{equation}
where $\bar u = t-\Tilde{r}_{\ast}$ and $\bar v = t+\Tilde{r}_{\ast}$ with $\Tilde{r}_\ast = \tanh^{-1}\LF H r_s \RF$.

\begin{figure}
    \centering
    \includegraphics[width=0.75\linewidth]{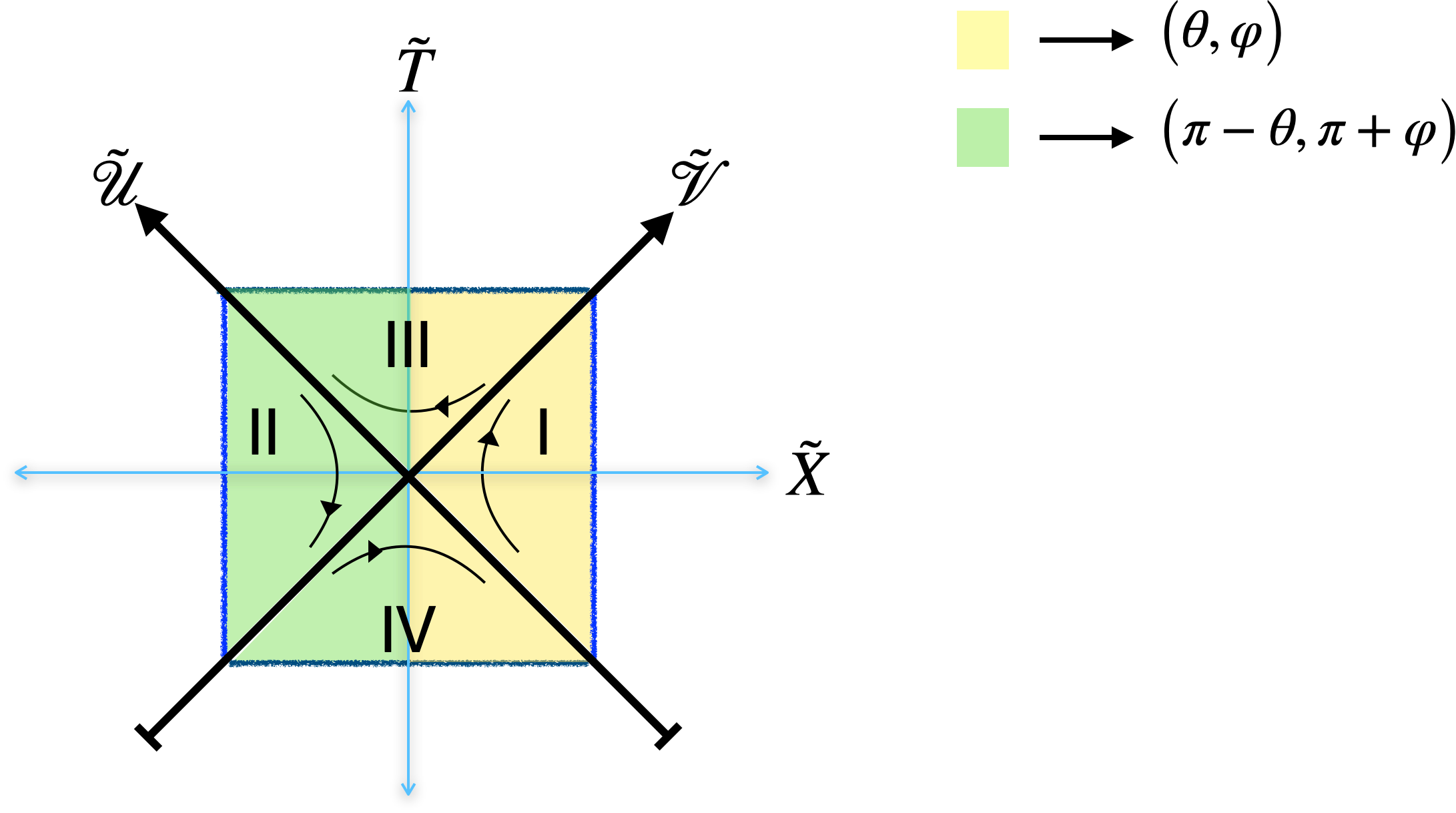}
    \caption{Here, we present the conformal diagram of de Sitter spacetime (in static coordinates) representing quantum states evolving according to dS DQFT. Every point in the yellow-shaded region and the green-shaded region correspond to angular coordinates $\LF \theta,\,\varphi \RF$, $\LF \pi-\theta,\,\pi+\varphi \RF$ respectively. The curvature of spacetime is constant everywhere in dS. According to dS DQFT quantum state in the entire region is a direct-sum of 4-components $ \vert \phi_{\rm total}\rangle = \frac{1}{\sqrt{2}} \LF \vert \phi_{I} \rangle \oplus \vert \phi_{II}\rangle \RF\oplus \frac{1}{\sqrt{2}} \LF \vert \phi_{III} \rangle\oplus \vert \phi_{IV}\rangle \RF$ which leads to unitarity in dS spacetime. The arrows in the figure indicate arrows of time in each region.}
    \label{fig:ds}
\end{figure}

According to the DQFT framework, Hilbert space or Fock space is always split whenever we encounter horizons or spacetime regions related by discrete spacetime transformations. Thus, a scalar field operator in static dS spacetime is split into direct-sum of 4 components 
\begin{equation}
    \hat{\phi}_{\rm static\,dS} = \frac{1}{\sqrt{2}}\LF \hat{\phi}_{I}  \oplus \hat{\phi}_{II}\RF \Big\vert_{r\gtrsim \frac{1}{H}} \oplus \frac{1}{\sqrt{2}}\LF \hat{\phi}_{III}  \oplus \hat{\phi}_{IV}\RF \Big\vert_{r\lesssim \frac{1}{H}} 
\end{equation}
The total Fock space now becomes direct-sum of 4 super selection sectors (which are the 4 regions of static de Sitter space Fig.~\ref{fig:ds}). 
\begin{equation}
    \Fc_{\rm static\,dS} = \LF \Fc_{I}\oplus  \Fc_{II}\RF \oplus \LF \Fc_{III}\oplus \Fc_{IV}\RF. 
\end{equation}
We apply the geometric superselection rule since all these 4-regions are related by discrete transformations (separated by horizons) \eqref{UVKruskaldS}. Moreover, the nature of spacetime outside $r\gtrsim \frac{1}{H}$ is static, whereas the interior $r\lesssim \frac{1}{H}$ is cosmological, which is analogous to the case of Schwarzschild BH metric \cite{Kumar:2024oxf,Kumar:2023hbj}. Thus, one cannot apply the same rules of quantum mechanics everywhere; once again, remember that time is a parameter in quantum theory. 

Consequently, the vacuum of static dS space is split as 
\begin{equation}
	\vert \Tilde{0}\rangle_{dS} = \LF\vert \Tilde{0}\rangle_{dSI} \oplus \vert \Tilde{0} \rangle_{dSII}\RF \oplus \LF\vert \Tilde{0}\rangle_{dSIII} \oplus \vert \Tilde{0} \rangle_{dSIV}\RF = \begin{pmatrix}
		\vert \Tilde{0} \rangle_{dSI} \\ 
		\vert \Tilde{0} \rangle_{dSII} \\
  \vert \Tilde{0} \rangle_{dSIII} \\ 
		\vert \Tilde{0} \rangle_{dSIV} 
	\end{pmatrix}\,.
\label{dsumFds1}
\end{equation}
In Fig.~\ref{fig:ds}, we depict the evolution of quantum states of the static dS space corresponding to all the regions defined in \eqref{UVKruskaldS}. A straightforward comparison we can make here is with Rindler spacetime presented in Appendix.~\ref{sec:Rindler}, which is based on \cite{Kumar:2024oxf}.

Similar to \eqref{puredS}, a maximally entangled pure state (at $r\gtrsim \frac{1}{H}$) becomes a direct-sum of pure states in the region I and region II of Fig.~\ref{fig:ds}

\begin{equation}
    \vert \Tilde{\psi}_{12}\rangle = \begin{pmatrix}
        \sum_{m,n} \Tilde{d}_{mn} \vert \phi_{I\,n1} \rangle \otimes  \vert \phi_{I\,n2}\rangle \\ 
        \sum_{m,n} \Tilde{d}_{mn} \vert \phi_{II\,m1} \rangle \otimes  \vert \phi_{II\,n2}\rangle
        \end{pmatrix}
    \label{LRent}
\end{equation}
where $\Tilde{d}_{mn} \neq \Tilde{d}_m\Tilde{d}_n$, $\vert \phi_1 \rangle = \sum_m \Tilde{d}_m\vert \phi_{m1}\rangle $ and $\vert \phi_2 \rangle = \sum_n \Tilde{d}_n\vert \phi_{n2}\rangle $. 
Therefore, Von Neumann entropy of density matrices in the sectorial Hilbert space of region I and II vanishes resulting in the purity of states in the entire dS spacetime.  
Thus, DQFT in dS promisingly complies with pure states evolving into pure states, bringing back the lost unitarity in the standard quantization methods. 

\subsection{Thermal spectrum and pure states}

This subsection highlights the important conclusion: The DQFT framework, by creating geometric superselection sectors based on discrete transformations, brings back the unitarity and observer complementarity. At the same time, the observer in dS spacetime does measure correlations \eqref{eqcorrdS} corresponding to the thermal spectrum of particles (quantum fields, to be precise) in a pure state (considering everything within the observer's horizon) as explained in the previous sub-sections. More details about this are presented in a companion paper \cite{Kumar:2024oxf} with analogies with unitary QFT in Rindler spacetime. It is worth noting that the thermal spectrum in a pure state is what is expected to have unitarity together with effects of curved spacetime, which is also very-well discussed in other investigations such as \cite{Kiefer:2001wn}.
In \cite{Kumar:2024oxf}, we also address further questions in relation to the implications of DQFT for the Reeh-Schlieder theorem known in SQFT \cite{Kumar:2024oxf}. 

\subsection{S-matrix and conformal diagram of de Sitter DQFT}

One important aim of QFTCS and quantum gravity is to describe scattering and compute amplitudes involving all gravitational and matter degrees of freedom. To achieve this, the first step is to understand semiclassically how to quantize a scalar field and describe the interactions perturbatively, neglecting back reactions. Unless we do this step (satisfying unitarity), the ultimate goal is always distant. As we discussed before, if we reach a sufficiently high energy limit, we should recover Minkowski QFT. It serves to verify if our approach to QFTCS is on the right track. To make progress, within this setup, we choose to ignore, for the moment all the popular claims about S-matrix in dS spacetime and/or dS cosmology \cite{Bousso:2004tv,Bousso:2002fq,Witten:2001kn,Balasubramanian:2020xqf}. Those claims stem from numerous classical notions before carrying quantization and are also based on hidden beliefs in the particularly popular frameworks of quantum gravity. 

In the sub-horizon limit $k^2\gg \frac{2}{\tau^2}$, we expect to recover the Minkowski limit. This should be understood in two ways. First of all, any mode is sub-horizon in the sufficient past, i.e., $a\to 0$ as $\tau\to \mp \infty$. As we noted before Sec.~\ref{sec:2to1}, the sign of $\tau$ does not determine spacetime since dS metric \eqref{conmetrc} is precisely time-symmetric \eqref{dSsym}. Secondly, at any given moment of dS expansion (i.e., at any value of $a$), we can always have sub-horizon modes. These are the two perspectives that are really important to understanding the problem of scattering in dS. The regular meaning of asymptotic states in QFT is bound to take a slightly different meaning in dS spacetime. 

Contrary to the popular expectations against the existence of a dS S-matrix \cite{Bousso:2004tv}, here we argue otherwise.  From the metric \eqref{dsmetric}, we can notice that in the time scales $\ll \frac{1}{\vert H\vert}$, the curvature effects can be negligible, and scattering can be completely treated like it is in Minkowski. In the limit where interactions involve high energy modes, within the time scales $\ll \frac{1}{\vert H\vert}$, the dS spacetime S-matrix should reduce to Minkowski spacetime S-matrix. This reasoning ensures the existence of an S-matrix in dS, and in fact, the same reasoning can be applied to any curved spacetime.

Imagine now that we have a particular set of modes that are all sufficiently sub-horizon (i.e., $k^2\gtrsim a^2H^2$, so we cannot completely neglect the background curvature) when they scatter. In the sufficient infinite past, we can treat them as modes that are deeper inside the horizon (i.e., $k^2\gg a^2H^2$ at which we completely neglect the background curvature effects and where they are very similar to Minkowski spacetime modes). At sufficient future, all the modes that scatter in the near sub-horizon become long wavelength modes, i.e., super-horizon $k^2\ll a^2H^2$, and act as free non-interacting states. After the modes become super-horizon, they are expected to evolve classically rather than quantum mechanically. Multifield inflation studies (See \cite{Gao:2008dt} and references therein for further details) have explored this extensively. 

As a toy model example, we consider two interacting KG fields given by 
\begin{equation}
	S_{\rm KGs} = \int \sqrt{-g} d^4x\LT \phi\square \phi + \chi\square \chi - g^2\chi \phi^2  \RT\,, 
	\label{intKGac}
\end{equation}
where $g$ is the coupling constant. As we discussed before, we quantize fields in dS after transforming them into harmonic oscillators by rescaling the fields $\phi\to a\phi,\, \chi\to a\chi$, which implies \\
\begin{equation}
	S^{\rm conformal}_{\rm KGs} = \int d\tau d^3x\LT \phi\square_\tau \phi + \chi\square_\tau \chi - g_{\rm eff}^2\chi \phi^2  \RT\,, 
	\label{dSSac}
\end{equation}
where $\square_\tau  = -\LF \pd_\tau^2+k^2 -\frac{2}{\tau^2} \RF$ and $g_{\rm eff} = \sqrt{a}g$ is the effective coupling constant which naturally becomes smaller as $a\to 0$. The action \eqref{dSSac}, in terms of the effective gravitational interaction $g_{\rm eff}$, is supposed to describe the scattering problem in dS at time scales $\Delta t \sim \frac{1}{\vert H\vert }$. If the scattering involves high energy modes, where all the modes remain deep inside the horizon all the way from beginning to end, this means, if $\Delta t \ll \frac{1}{\vert H
\vert}$,  then for practical purposes, the background curvature effects can be dropped through the rescaling  $a\to 1$ and $g_{\rm eff}\to g$ and we can follow DQFT in Minkowski spacetime.

\begin{figure}[ht!]
	\centering
	\includegraphics[width=0.6\linewidth]{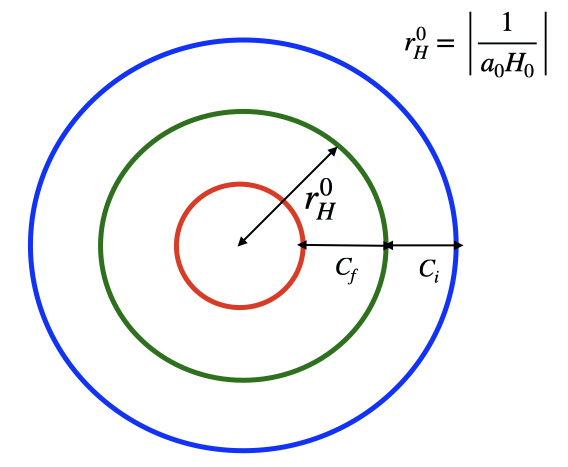}
	\caption{We depict the scattering process schematic picture in expanding dS. Here, $r_H^0$ denotes the scale at which the modes start to interact and undergo scattering. In the sufficient past (i.e., when the comoving horizon was $r_H = r_H^0+C_i$ where $C_i$ is some arbitrary constant radius), we assume the modes are sufficiently sub-horizon and non-interacting. Whereas in the sufficient future  (i.e., when the comoving horizon will be $r_H = r_H^0-C_f$ where $C_f$ is some arbitrary constant radius), we assume that the modes, after scattering, exit beyond radius $r_H^0$ and become classically evolving fluctuations.  }
	\label{fig:dss-matrix}
\end{figure}

We present our expected picture for dS scattering in Fig.~\ref{fig:dss-matrix}. 
Action \eqref{dSSac} seems to describe two interacting scalar fields in flat spacetime, with time ($\tau$) dependent masses and coupling constant $g_{\rm eff}$.  
At this point, and taking into account Fig.~\ref{fig:dss-matrix}, we can now write a dS DQFT S-matrix in analogy with \eqref{dQFTSM} and \eqref{DQFTSM12} as 
\begin{equation}
	\begin{aligned}
		S_{dS} = S_{dSI} \oplus S_{dSII} 
		\end{aligned}
	\label{dSSMatrix}
\end{equation}
where 
\begin{equation}
	S_{dSI} = T_{dSI} \Bigg\{  e^{-i\int_{-r_H^0-C_i}^{-r_H^0+C_f} H_{int} \,\, d\tau } \Bigg \},\quad S_{dSII}  = T_{dSII} \Bigg\{ e^{i\int_{r_H^0+C_i}^{r_H^0-C_f} H_{int } \,\,  d\tau} \Bigg \}
	\label{DQFTDS12}
\end{equation}
with $T_{dSI},\, T_{dSII}$ representing the time orderings attached to the respective sectorial Fock spaces $\Fc_{dSI},\,\Fc_{dSII}$ \eqref{dSFock} arrows of time. 
We can notice that, in the limits $C_i,\, C_f\to \infty$, we recover \eqref{DQFTSM12} and show the consistency of DQFT S-matrix limiting behavior in dS with the corresponding short distance Minkowski case. 

\begin{figure}[ht!]
	\centering
	\includegraphics[width=0.65\linewidth]{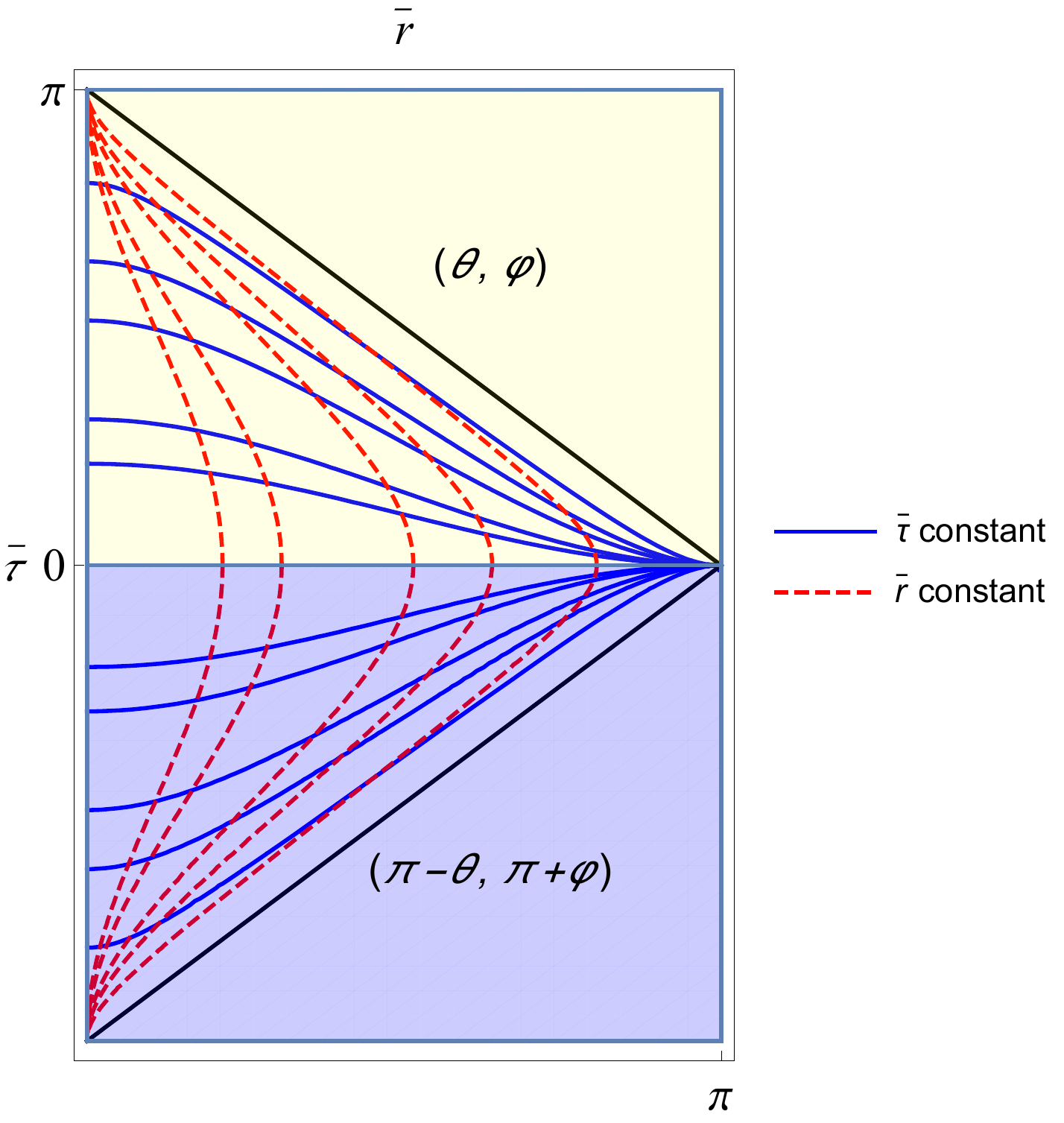}
	\caption{This picture is a conformal diagram representing DQFT in (expanding) flat FLRW dS in terms of parametric time $\bar{\tau}$ and the compactified radial coordinate $\bar{r}$. {The shaded quadrants, in yellow and blue, represent spatial regions related by parity, where conformal time is positive in the yellow region and negative in the blue region.} A field operator in dS DQFT is represented by the direct-sum of field operators populating the yellow and blue regions following \eqref{disumdS}.   }
	\label{fig:dsjoined}
\end{figure}

In the context of Minkowski space DQFT, we provided a pictorial representation of a conformal diagram in Fig.~\ref{fig:minkowski-np}. We can replicate this procedure and draw a conformal diagram in the context of (expanding) dS spacetime,  which reflects the interpretation of dS DQFT. We present Fig.~\ref{fig:dsjoined} in terms of the compactified parametric time $\bar{\tau}$ and the radial coordinate $\bar{r}$ \cite{Mukhanov:2005sc} 
\begin{equation}
	\bar{\tau} =\dfrac{\sin\left( \eta \right) }{\cos\left( \eta \right) + \cos\left(\tilde \chi \right)} ,\quad \bar{r} = \dfrac{\sin\left( \tilde\chi \right) }{\cos\left( \eta \right) + \cos\left(\tilde \chi \right)}
\end{equation}

Finally, we would like to comment a bit further on the significance of DQFT S-matrix definition in dS \eqref{dSSMatrix}. As we can clearly verify,  \eqref{dSSMatrix} is invariant under $\Pc\Tc$ as far as the interaction Hamiltonian obeys the symmetry. Furthermore, we can now understand that the meaning of asymptotic states that we define in QFT are slightly different in curved spacetime, where the horizon effects must be seriously taken into account. However, it seems not an impossible task to make sense of QFTCS as it is expected from the most recent seminal papers by Giddings and Witten \cite{Giddings:2022jda,Witten:2021jzq}. We anticipate that our approach to QFTCS could contribute to understanding the subject further. The exact computation of a scattering process in dS DQFT would be a future subject of our investigation. {In this regard, we aim to generalize the results of scattering amplitudes in dS \cite{Ferrero:2022tqn,Ferrero:2021lhd} in our DQFT formalism, which is unitary.}
In the next section, we will discuss and propose views about the implications of our study to the quantum gravity research field. 

 \subsection{Comments about de Sitter QFT in closed or global de Sitter}

As discussed earlier, dS spacetime can be expressed in several coordinate frames due to its symmetries. When we express the dS metric in a particular set of coordinates, we automatically fix our imaginary observer. In the previous section, we discussed dS in flat FLRW, the most relevant one for applications in inflationary cosmology. However, closed dS metric \eqref{cloFLRW} or its most often called the global dS spacetime \cite{Hartman:2017} has been the popular subject of investigation from a theoretical point of view \cite{Bousso:2002fq}. The imaginary observer, in closed dS spacetime, experiences the Universe evolving from contraction to expansion (defined by treating the "scale factor" as a clock, not the coordinate time). To implement direct-sum quantization, we need a notion of time reversal operation without changing the character of the Universe (i.e., either expanding or contracting), which is given by \eqref{clldS}. Such as, in the flat FLRW dS case \eqref{disumdS}, we can analogously expand the scalar field operator as the direct-sum of two components in closed dS. However, the difference is that the (direct-sum) vacuum of closed dS would differ because the mode functions are also different. Physically, during contraction, we will have the modes entering the horizon, and during the expansion phase, we will have the modes exiting the horizon, like in the flat FLRW (expanding) dS case. Like in flat FLRW, we can analogously have unitarity and observer complementarity in closed dS with a direct-sum Fock space structure. Defining the scattering problem in closed dS is actually contextual because any scattering process can happen during i) the contraction phase, ii)  the expansion phase, and iii) the transition from contraction to expansion. S-matrix formulation in closed dS was addressed in the past \cite{Srednicki:2007qs}, but such framework only discusses case (iii), with pre-scattering states in the contracting phase and post-scattering states in the expanding phase. We note that this is a very particular scenario for a scattering problem. Generically, scattering can happen at any stage of the Universe's evolution. Therefore, it is vital to study cases (i), (ii), and (iii) very carefully. 
Having mentioned this, we indeed defer the detailed studies of closed dS to future investigations.

\section{Possible implications of direct-sum QFTCS to quantum gravity}
\label{sec:DQFT-QG}
There are two points of view about dS within the quantum gravity research community. The first is to formulate a quantum gravity theory whose classical or semiclassical limit is dS \cite{Susskind:2021omt,Aalsma:2020aib,Balasubramanian:2001rb}. This direction of research is an extremely arduous one. There are different proposals on how one can achieve this in the formulation of quantum gravity \cite{Hartman:2020khs,Shaghoulian:2021cef,Shaghoulian:2022fop,Balasubramanian:2021wgd,Chakraborty:2023yed,Maniccia:2023cgv}. 
 Furthermore, the understanding of dS is most often dominated by the holography conjecture approach to quantum gravity  \cite{Strominger:2001pn,Bousso:2002fq,Fabinger:2003gp,Geng:2020kxh} where one sees the appearance of two asymptotic conformal field theories (CFTs) corresponding to the bulk description of gravity. 
 Although the present work is not fully within the context of quantum gravity, we propose that one should explore the DQFT formulation in a holographic version of quantum gravity.   
We think DQFT in curved spacetime might also be useful in making a new attempt to realize dS in string theory. 

We would also like to comment on the difference between DQFT in (expanding) FLRW flat dS and the elliptic view of dS, which is studied from the holographic point of view \cite{Parikh_2003,Parikh_2005}. First of all, elliptic dS is a proposal dating back to Schr\"{o}dinger \cite{Schrodinger1956}, which is based on the classical identification of two expanding branches of dS in the ellipsoid that is embedded in 5D Minkowski. In \cite{Parikh_2003,Parikh_2005}, S-matrix is proposed for dS, which is fully based on this classical connection of spacetime regions through antipodal identification. Such a framework becomes problematic once we depart from dS spacetime, where the antipodal identification leads to closed time-like curves  \cite{Aguirre_2003}. In the DQFT in curved spacetime approach, we do not invoke any classical identification of spacetime. Rather, we construct a description of the quantum field with two arrows of time using the geometric superselection Fock spaces corresponding to parity conjugate regions of physical space. Therefore, this framework will have interesting implications for the quantum gravity program in dS  \cite{Witten:2001kn}.  
 
The second point of view about dS, within quantum gravity research, is the bottom-up approach, which aims to formulate a robust QFTCS before stepping into the fully unknown arena of quantum gravity. This is exactly the point of view taken in this paper. We would like to acknowledge several studies that have explored the stability of dS spacetime via quantum loop corrections due to matter fields \cite{Akhmedov:2013vka,Green:2022ovz}, which have given us a way to build effective field theories (EFTs) of gravity for dS spacetime. 
We think our dS DQFT proposal will further shed light on building a unitary QFTCS and, eventually, allow us to compute observables such as scattering amplitudes, which can lead to interesting implications in the context of cosmological applications beyond current studies \cite{Ferrero:2021lhd,Ferrero:2022tqn,Arkani-Hamed:2015bza,Baumann:2022jpr}.

\section{Conclusions and Outlook}
\label{sec:conc}

QFTCS constitutes an important field of study before forming a consistent framework for quantum gravity. The main difficulty for QFTCS (in the context of curved spacetimes with horizons) is the loss of unitarity. However, we must stress that this difficulty is expected and, in fact, is why GR and quantum mechanics are often said to be incompatible. QFTCS is an inevitable obstacle that needs to be overcome before tackling the unknown Planck scale physics, unification of fundamental forces, or the issue of GR non-renormalizability. 
Surely, we do not ignore several decades of developments in quantum gravity. We intend to highlight the unavoidable importance of QFTCS, which surely gives us better foundations for bigger challenges such as quantum gravity. Let us emphasize that we are speaking about QFTCS in a totally semiclassical sense, i.e., just up to the KG field quantization in curved spacetime. In this paper, we highlighted that modeling de Sitter spacetime to represent an expanding Universe (growing scale factor in flat FLRW coordinates) is compatible with two arrows of time (as explained in Sec.~\ref{sec:2to1}, see \eqref{expcon}), which is usually understood as two Universes. Given that we only observe one Universe, choosing an arrow of time by hand breaks the symmetry of spacetime and leaves the possibility of another Universe. Our focus is to write a quantum theory in dS spacetime that takes into account the two arrows of time and achieves unitarity. 

We start with background (discrete) symmetries of the spacetime and prepare the quantum field operators to obey the discrete symmetries, which we achieve by geometric superselection rules (or direct-sum structure of Fock space). 
This has led us to obtain unitarity (in the sense that pure states should not evolve to mixed states within the horizon), ultimately making the right platform for a consistent QFTCS. In contrast to the major opinions about the impossibility of constructing unitary QFTCS with an S-matrix\footnote{S-matrix understanding is needed to understand any scattering process in curved spacetime. Scattering in curved spacetime makes sense. We know particles (quantum fields) exist and they can scatter whether or not the background spacetime is curved.  Moreover, any scattering in curved spacetime must look like the one in Minkowski for the wavenumbers exceeding the background spacetime curvatures. In fact, in the context of early Universe cosmology, imprints of particle scatterings in the cosmological correlations is an active field of research nowadays \cite{Arkani-Hamed:2015bza,Baumann:2022jpr}. However, these studies avoid discussions of S-matrix and unitarity (pure states to pure states), by following so-called cosmological bootstrap techniques based on wave function of Universe in the asymptotic (temporal) limit. Nevertheless, the scattering problem in curved spacetime should have a resolution; it is just that we do not yet clearly know how to formulate a robust QFTCS. It is often stated that the S-matrix is not observable in curved spacetime. This is again associated with the fact that the standard QFTCS does not give us a way to define {\it in} and {\it out} states (strictly speaking, pre-scattering and post-scattering states), preserving unitarity. 
	 In this paper, we define quantum states in dS spacetime as having a well-defined scattering within the (co-moving) horizon. We will address the calculation of scattering amplitudes in detail in future investigations. The meaning of S-matrix in BH physics is the subject we explored \cite{Kumar:2023hbj}.}, we start with fundamental questions about the concept of time in quantum theory (where it is just a parameter). We started our investigation by revising the quantum theory and its construction based on the arrow of time and the definition of the positive energy state. We rethink it is possible to make several non-trivial observations regarding how we perform quantization in Minkowski's background, which actually follows from our conceived notion of a time arrow. This fact shows us how important discrete symmetries are in quantum theory, and this might constitute a good starting point for a rethink of QFTCS.  

{In this paper, we developed a direct-sum quantum theory in which a single quantum state is expressed as the direct-sum of two components that belong to sectorial Hilbert spaces representing parity conjugate regions of physical space. In this picture, we incorporate two arrows of time to represent a single positive energy state.} The DQFT approach we established opens a new door for the meaning of time in QFT, and it is certainly useful in tackling unitarity issues in curved spacetime. In DQFT, Fock space is constructed with geometric superselection rules dictated by discrete spacetime transformations. We obtain unitarity and observer complementarity when we implement such a construction for QFTCS, especially in dS spacetime. 
This construction allows an (imaginary) observer in curved spacetime to construct a complete set of states within his/her horizon and reconstruct information beyond the horizon by the knowledge of (pure) states within the horizon. This is due to the fact that in DQFT in curved spacetime, the horizon (quantum mechanically) acts like a "mirror" associated with respective discrete spacetime operations.
Furthermore, we also laid out some novel thoughts for defining the scattering problem (S-matrix) in dS. 
Our approach to QFTCS is theoretical but has been successfully applied to explain CMB anomalies with a new formulation of inflationary quantum fluctuations  \cite{Gaztanaga:2024vtr,Gaztanaga:2024whs} and provided new predictions for primordial gravitational waves \cite{Kumar:2022zff}. A future extension of our work would be to understand and fully compute scattering amplitudes in DQFT in dS spacetime. There is still a long way to a definitive QFTCS, and we expect the subject to resolve many conundrums and surprising revelations further. In \cite{Kumar:2023hbj}, we addressed another challenging spacetime in QFTCS: Schwarzchild BH.  

Our final remark is,  
Unitarity in QFTCS is an important barrier we must first cross to achieve a consistent quantum gravity, and we sincerely hope our present investigation paves the way for interesting explorations in merging the quantum \& gravity.

\acknowledgments

KSK acknowledges the support from JSPS and KAKENHI Grant-in-Aid for Scientific Research No. JP20F20320 and No. JP21H00069, and thank Mainz U. and Beira Interior U. for hospitality where part of the work has been carried out. KSK would like to thank the Royal Society support in the name of Newton International Fellowship. 
 J. Marto is supported by the grant UIDB/MAT/00212/2020  and COST action 23130. We would like to thank Chris Ripken for the initial collaboration on the project and giving us a lot of insights into the algebraic QFT.  We thank Gerard 't Hooft for truly inspiring us over the years with his impactful insights into the subject and also we thank him for the very useful comments and discussions during the preparation of this manuscript. We thank Yashar Akrami, Norma G. Sanchez, Masahide Yamaguchi, Alexei A. Starobinsky, Luca Buoninfante, Francesco Di Filippo, Yasha Neiman, Paolo Gondolo, Martin Reuter, Eiichiro Komatsu, Paulo V. Moniz, Gia D'vali, David Wands, Mathew Hull, and Enrique Gazta\~naga for useful discussions. KSK would like to especially thank Sivasudhan Ratnachalam for positive encouragement, friendship, and useful discussions on quantum mechanics.

\appendix

\section{DQFT in Rindler spacetime}

\label{sec:Rindler}

In this section, we review the unitary formulation of QFT (i.e., DQFT) in Rindler spacetime formulated recently in \cite{Kumar:2024oxf}. Elements of this section complement the understanding of DQFT in dS we presented in Sec.~\ref{sec6DQFTDS}.

\begin{figure}
    \centering
    \includegraphics[width=0.5\linewidth]{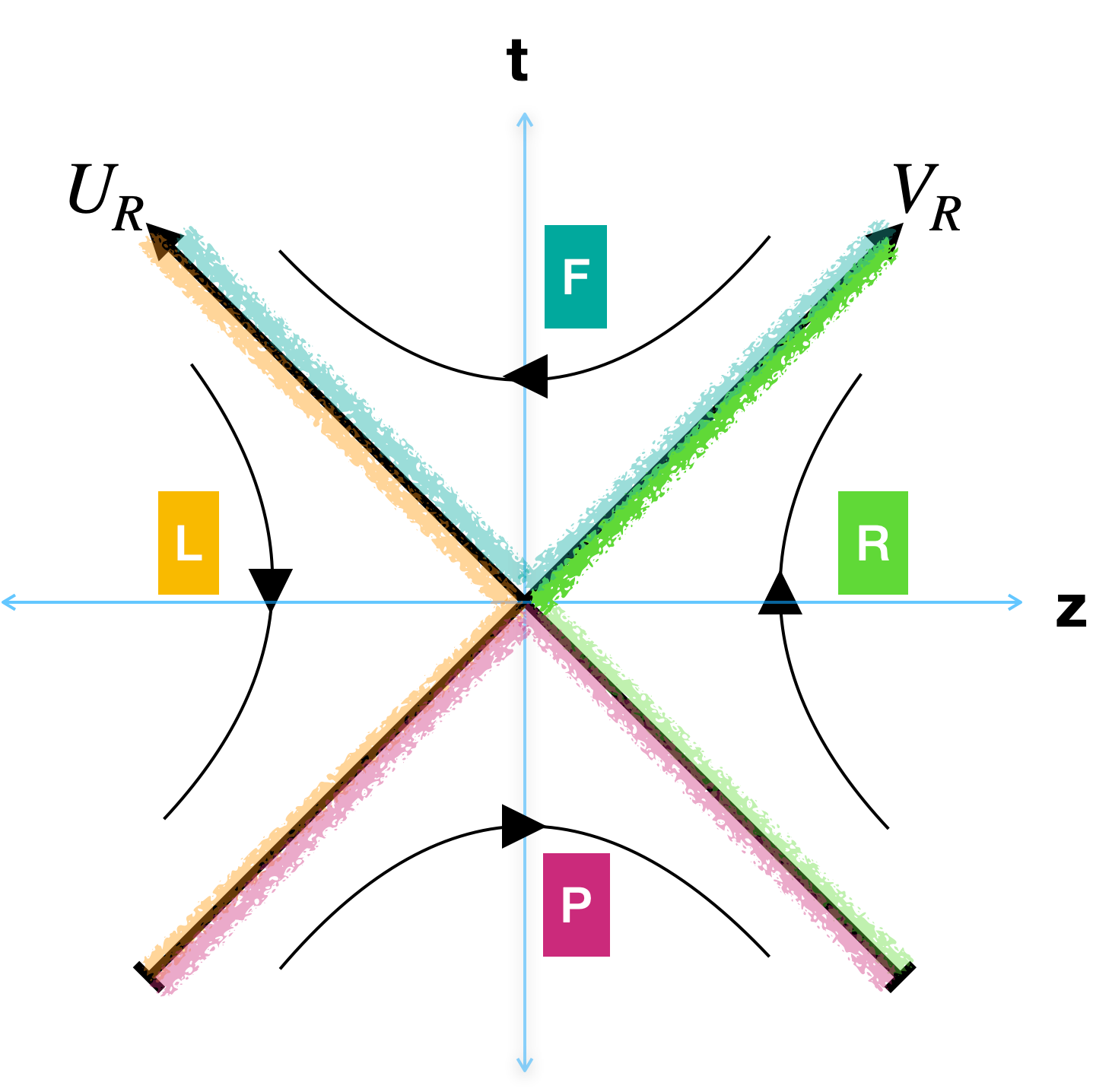}
    \caption{In this picture, we depict the Rindler spacetime with the Left, Right ($z^2\gtrsim t^2$), and Future, Past ($t^2\gtrsim z^2$) regions in Minkowski plane $\LF t,\,z \RF$. The curved lines with arrows in the Left and Right regions depict a constant acceleration $ae^{-a\xi}$ in the direction $\eta: \infty \to -\infty$ and $\eta: -\infty \to \infty$ respectively. Future and Past Rindler wedges are the degenerate Kasner Universes where the arrows are changing $z: \pm\infty \to \pm\infty$, which means $\eta: \mp\infty \to \pm\infty$. The fuzzy colored lines indicate Rindler Horizons for Left (Yellow), Right (Green), Future (Cyan), and Past (Pink).}
    \label{fig:RindlerST}
\end{figure}

Rindler spacetime is obtained by coordinate transformation of Minkowski spacetime (let us consider 1+1 dimensions for simplicity) 
\begin{equation}
    ds^2 = -dt^2+ dz^2
\end{equation}
\begin{equation}
\begin{aligned}
 z^2-t^2 & = \frac{1}{a^2} e^{2a\xi} \implies \begin{cases}
     z  = \frac{1}{a} e^{a\xi} \cosh{a\eta},\quad  t= \frac{1}{a} e^{a\xi} \sinh{a\eta} \quad \LF \textrm{Right Rindler} \RF \\   z  = -\frac{1}{a} e^{a\xi} \cosh{a\eta},\quad  t= \frac{1}{a} e^{a\xi}\sinh{a\eta} \quad \LF \textrm{Left Rindler} \RF
 \end{cases} \\ &\implies \boxed{ds^2 = e^{2a\xi}\LF -d\eta^2+d\xi^2\RF} \\   
  t^2-z^2 & = \frac{1}{a^2} e^{2a\eta} \implies \begin{cases}
     t  = \frac{1}{a} e^{a\eta} \cosh{a\xi},\quad  z= \frac{1}{a} e^{a\eta} \sinh{a\xi}\quad \LF \textrm{Future Kasner} \RF  \\  t  = -\frac{1}{a} e^{a\eta} \cosh{a\xi},\quad  z= \frac{1}{a} e^{a\eta} \sinh{a\xi}\quad \LF \textrm{Past Kasner} \RF 
 \end{cases} \\ & \implies \boxed{ds^2 = e^{2a\eta}\LF -d\eta^2+d\xi^2\RF}
    \end{aligned}
    \label{Robserco}
\end{equation}
The description of this spacetime is presented pictorially in Fig.~\ref{fig:RindlerST}.

We can rewrite the whole Rindler spacetime (with all regions) in a coordinate system defined by 
\begin{equation}
    \begin{aligned}
        U_R &= -\frac{1}{a}e^{-au}<0,\quad &&V_R= \frac{1}{a}e^{av}>0\quad &&(\rm Right\, Rindler)\\
        U_R&= \frac{1}{a}e^{-au}>0,\quad &&V_R= -\frac{1}{a}e^{av}<0\quad &&(\rm Left\, Rindler) \\
         U_R&= \frac{1}{a}e^{-au}>0,\quad &&V_R= \frac{1}{a}e^{av}>0\quad &&(\rm Future\, Kasner) \\
          U_R&= -\frac{1}{a}e^{-au}<0,\quad &&V_R= -\frac{1}{a}e^{av}<0\quad &&(\rm Past\, Kasner)
    \end{aligned}
    \label{UVcoord}
\end{equation}
where 
\begin{equation}
\begin{aligned}
    u&= \eta-\xi,\quad v=\eta+\xi \\ 
    U_R&= t-z,\quad V_R=t+z
    \end{aligned}
\end{equation}
The coordinates \eqref{UVcoord} define the Left, Right, Future, and Past Rindler regions separated by the horizons carried through discrete transformations on $\LF U_R,\, V_R \RF$. 
These are analogous to the Kruskal coordinates in dS \eqref{UVKruskaldS} and black hole spacetime \cite{Kumar:2023hbj}.  

According to DQFT in Rindler spacetime \cite{Kumar:2024oxf}, a (maximally) entangled pure state (in $z^2-t^2\gtrsim 0$) becomes a direct-sum of two pure state components in Left Rindler and Right Rindler regions 
\begin{equation}
    \vert \psi_{LR}\rangle = \frac{1}{\sqrt{2}}\begin{pmatrix}
        \vert \phi_{R1} \rangle \otimes \vert \phi_{R2} \rangle \\ 
        \vert \phi_{L1} \rangle \otimes \vert \phi_{L2} \rangle 
    \end{pmatrix} 
    \label{LRent1} 
\end{equation}  
A similar conclusion can be derived for the Future and Past regions. Thus, the Von Neumann entropy of the pure state components in the each region vanishes which leads to unitary description of quantum fields (or states) for all the observers. Since all the regions are related by discrete transformations, observing a pure state an observer in any region can reconstruct information beyond the Rindler horizon. 
We refer to \cite{Kumar:2024oxf} for further details.

\bibliographystyle{utphys}
\bibliography{ref.bib}

\end{document}